\documentclass[11pt,a4paper]{article}
\pdfoutput=1
\usepackage{mystyle}

\usepackage{graphicx}
\usepackage{braket}
\usepackage{latexsym}
\usepackage{amsmath}
\usepackage{mathrsfs}
\usepackage{amssymb}	

\usepackage{slashed}
\usepackage{tabularx}
\usepackage{tabu}
\usepackage{cancel}

\usepackage{multirow}
\usepackage{booktabs}
\usepackage{lipsum}

\usepackage{tikz}
\usetikzlibrary{arrows,decorations.pathmorphing,backgrounds,positioning,fit,petri,automata,calendar,mindmap,
decorations.markings,calc}

\newcommand{\tev}{{\rm TeV}}
\newcommand{\gev}{{\rm GeV}}

\newcommand{\nn}{\nonumber}

\newcommand{\be}{\begin{equation}}
\newcommand{\ee}{\end{equation}}
\newcommand{\bea}{\begin{eqnarray}}
\newcommand{\eea}{\end{eqnarray}}

\newcommand{\bfr}{\begin{mdframed}[backgroundcolor=gray!20] }
\newcommand{\efr}{\end{mdframed}}

\newcommand{\gY}{g'}
\newcommand{\gL}{g}

\def\lra#1{\overset{\text{\hspace{.2em}\raisebox{-.1em}{\footnotesize$\leftrightarrow$}}}{#1}}

\newcount\hour \newcount\minute
\hour=\time \divide \hour by 60
\minute=\time
\begin{document}


\title{A global view on the Higgs self-coupling}

\author[a]{S.~Di Vita,}
\author[a,b,1]{C.~Grojean,\note{On leave from Instituci\'o Catalana de Recerca i Estudis Avan\c cats, 08010 Barcelona, Spain}}
\author[c]{G.~Panico,}
\author[a,c]{M.~Riembau,}
\author[a,c]{T.~Vantalon}

\affiliation[a]{DESY, Notkestra{\ss}e 85, D-22607 Hamburg, Germany}
\affiliation[b]{Institut f\"ur Physik, Humboldt-Universit\"at zu Berlin, D-12489 Berlin, Germany}
\affiliation[c]{IFAE, Barcelona Institute of Science and Technology (BIST) Campus UAB, E-08193 Bellaterra, Spain}

\emailAdd{stefano.divita@desy.de}
\emailAdd{christophe.grojean@desy.de}
\emailAdd{gpanico@ifae.es}
\emailAdd{marc.riembau@desy.de}
\emailAdd{tvantalon@ifae.es}

\abstract{The Higgs self-coupling is notoriously intangible at the
  LHC. It was recently proposed to probe the trilinear Higgs interaction
   through its radiative corrections to single-Higgs processes.
  This approach however requires to disentangle these effects from those associated to
  deviations of other Higgs-couplings to fermions and gauge bosons. 
  We show that a global fit exploiting only single-Higgs inclusive data suffers
  from degeneracies that prevent one from extracting robust
  bounds on each individual coupling.  We show how the inclusion of
  double-Higgs production via gluon fusion, and the use of differential
  measurements in the associated single-Higgs production channels
  $WH, ZH$ and $t\bar{t}H$, can help to overcome the deficiencies of a
  global Higgs-couplings fit. In particular, we bound the variations of
  the Higgs trilinear self-coupling  relative to its SM value to the interval $[0.1, 2.3]$ at
  $68\%$ confidence level at the high-luminosity LHC, and we discuss the robustness of our results
  against various assumptions on the experimental uncertainties and
  the underlying new physics dynamics. We also study how to obtain a
  parametrically enhanced deviation of the Higgs self-couplings and we
  estimate how large this deviation can be in a self-consistent
  effective field theory framework.  }

\preprint{\texttt{DESY 17-044}}

\maketitle

\section{Introduction}

The exploration of the energy frontier is happening now: 2016 has been a record year for the LHC that accumulated an unprecedented amount of luminosity at the highest energy of 13\,TeV~\cite{Bordry}.
In the absence of the long awaited bump revealing the first putative supersymmetric partner needed to stabilize the weak scale, to unify the fundamental interactions, to account for dark matter among other things, it should not be forgotten that the LHC is more than a discovery machine.  It can also be used for precision measurements giving an extra handle to reveal the existence of new physics. In this roadmap, the acclaimed Higgs boson plays a central role: with all its couplings uniquely predicted in the Standard Model (SM), it is the new metronome that can serve to quantify the accuracy of the SM description of our world. Major efforts have been devoted first to provide consistent theoretical frameworks to deform the SM Higgs couplings in a way as model independent as possible, and second to pin down or at least bound these deformations using the experimental data (see for instance refs.~\cite{Olive:2016xmw, deFlorian:2016spz}). A quantity of particular interest but notoriously intangible is the Higgs cubic self-interaction. It is even often said that the value of this coupling is a key to check the electroweak symmetry breaking. Indeed, the SM Higgs potential, is given after breaking by
\begin{gather}
\mathcal{L} \supset  -\frac{m_h^2}{2} h^2 - \lambda_3^{SM} v h^3 -\lambda_4^{SM} h^4,\\
\lambda_3^{SM}= \frac{m_h^2}{2 v^2}, \qquad \lambda_4^{SM}= \frac{m_h^2}{8 v^2},
\end{gather}
where the Higgs vacuum expectation value (VEV) $v \simeq 246$~GeV can be related to the Fermi constant and measured in muon decay, and $m_h$ is precisely determined by fitting a bump in the di-photon and the four-lepton decay channels. And measuring $\lambda_3$ is a good way to check that electroweak symmetry breaking (EWSB) follows from a simple Ginzburg--Landau $\phi^4$ potential.
Moreover many different Beyond the Standard Model (BSM) scenarios allow large deviations for the Higgs self-couplings (see for instance ref.~\cite{Olive:2016xmw}), and measuring $\lambda_3$ could be a way to probe the existence of new physics.

Until recently, the main approach to constrain the Higgs cubic self-coupling (the quartic is likely to remain impalpable before long) was relying on the measurement of the  double Higgs  production rate that directly depends, at leading order (LO), on the value of $\lambda_3$. This measurement is, however, complicated by the low cross section as well as the small decay fractions in the channels that can compete against the ever annoying dominant QCD background. And the sensitivity remains frustrating low. A few years ago, ref.~\cite{McCullough:2013rea}  proposed to measure/constrain the Higgs cubic self-coupling at $e^+e^-$ colliders via the quantum corrections it induces in single Higgs  channels. Recently this idea has been revisited at hadronic machines by refs.~\cite{Gorbahn:2016uoy,Degrassi:2016wml,Bizon:2016wgr}, which concluded that potentially this approach could be complementary if not competitive or even superior to the traditional strategy.
This idea has also been further extended to bound the Higgs self-coupling deviations using EW  precision measurements~\cite{Degrassi:2017ucl, Kribs:2017znd} with the conclusion again that competitive results can be derived.
Such dramatically optimistic conclusions deserve to be scrutinized and disputed.
First it should be noted that those analyses look at scenarios where only the Higgs self-coupling deviates from the SM.
After understanding which particular UV dynamics would fulfill this assumption, one should question the robustness of their conclusions under less restrictive hypotheses. A corollary question is also  to understand to which extend the traditional and simple fits of the single Higgs couplings, that were neglecting the effects of the Higgs trilinear, could get distorted. Truly model-independent bounds on the Higgs couplings cannot be obtained. It is of the uttermost importance to be alerted on the sometimes hidden dynamical assumptions sustaining the bounds derived from a particular fit. And be aware of the classes of models these bounds safely apply to.

Even in models where the Higgs self-coupling receives a correction parametrically enhanced compared to the deviations of the other Higgs couplings, a careful multi-dimensional analysis is in order. Indeed, even loop suppressed deviations to couplings to gauge bosons or fermions will affect at LO single Higgs processes, whereas the Higgs trilinear coupling enters at next-to-leading order (NLO). So both effects can typically be of the same order. And to set bounds on each coupling deviation, a complete and global fit is needed. This is the main question we address in this paper. We first notice that a fit to the inclusive single Higgs observables alone suffers from a blind direction and that it is not possible to bound individually each of the coupling nor to extract any information on the Higgs trilinear interaction. We advocate that extra observables are needed to resolve this degeneracy. We first focus our attention on the benefit of including information on the double Higgs production channels. We then explore the use of differential kinematic distributions in single Higgs processes and we conclude that it is a promising idea that however requires a realistic and detailed estimate of the systematic uncertainties. However, we caution that in  scenarios that produce enhanced deviations in the Higgs self-couplings, one should also pay particular attention to the constraints imposed by electroweak precision measurements that could be,  for Higgs physics,  far less restrictive than in generic BSM models, requiring an even more global fit of Higgs and EW data together.

Our paper is organized as follows. In section~\ref{sec:eft_parametrization}, we introduce the notation conventions and the effective field theory (EFT) parametrization we will use to capture the deformations of the Higgs couplings. We spend some efforts to identify and characterize possible UV dynamics that could give rise in a self-consistent way to large corrections to the Higgs trilinear coupling. In section~\ref{sec:inclusive}, we perform a global analysis using inclusive single-Higgs observables. We show that it is plagued with an exact flat direction and we discuss how this degeneracy affects the traditional determination of the single Higgs couplings. In section~\ref{sec:double_Higgs}, we show how double-Higgs production could rescue a Higgs-couplings global fit. In section~\ref{sec:differential}, we speculate that single-Higgs differential observables could also help constraining the Higgs couplings in a global way and we discuss the robustness of the results of our proposed fit against several implicit assumptions used. Section~\ref{sec:conclusions} presents our conclusions while two appendices contain some technical details of our EFT analysis. In all our projections, we consider $3$\,ab$^{-1}$ of cumulated luminosity collected at $14$\,TeV and we rely on conservative estimates of the systematic uncertainties in the various Higgs production and decay channels reported in ECFA
studies~\cite{ATLAS:projections,ATL-PHYS-PUB-2016-008}.

\section{The effective parametrization}\label{sec:eft_parametrization}

\subsection{Higgs primary couplings}

In a large class of scenarios, if a sizable gap is present between the SM states and the mass scale of the BSM dynamics,
the new-physics effects can be conveniently encapsulated into an EFT framework.
The EFT operators can be organized according to their canonical dimension, thus expanding the effective Lagrangian
into a series
\begin{equation}
\mathcal{L} = \mathcal{L}_{\textsc{sm}} + \sum_{i} \frac{c^{(6)}_i}{\Lambda^2} \mathcal{O}^{(6)}_i
+ \sum_{i} \frac{c^{(8)}_i}{\Lambda^4} \mathcal{O}^{(8)}_i + \cdots\,,
\end{equation}
where $\mathcal{L}_{\textsc{sm}}$ is the SM Lagrangian, $\mathcal{O}_i^{(D)}$ denote operators of dimension
$D$ and $\Lambda$ is the SM cut-off, i.e.~the scale at which
the new dynamics is present.\footnote{In the above expansion we neglected operators with odd energy dimension
since they violate lepton number conservation (for $D = 5$) and $B - L$ invariance (for all odd $D$). These effects are constrained
to be extremely small and do not play any role in our analysis.}

The leading new-physics effects are usually associated
with EFT operators with the lowest dimensionality, namely the dimension-6 ones.
In the following we restrict our attention to these operators and neglect higher-order effects.
To further simplify our analysis we also assume that the new physics is CP-preserving
and flavor universal. With these restrictions we are left with $10$ independent operators that affect Higgs physics
at leading order and have not been tested below the $\%$ accuracy in existing precision measurements~\cite{Falkowski:2015fla}.\footnote{The
assumption of flavor universality is not crucial for our analysis. It is only introduced to restrict the EFT analysis to the
operators that can only be tested in Higgs physics. The same can be done in several other flavor scenarios,
as for instance minimal flavor violation and anarchic partial compositeness.}

Before discussing our operator basis, it is important to mention that a much larger set of dimension-6 operators
could in principle be relevant for Higgs physics. A first class of these operators include deformations of the SM Lagrangian
involving the light SM fermions. They correct at tree level the Higgs processes but also affect observables
not involving the Higgs. Therefore most of them have already been tested with good precision in EW measurements.
A second set of dimension-6 operators involve the top quark and are typically much less constrained. However they affect
Higgs physics only at loop level, thus their effects are usually not very large. We postpone a more detailed discussion to section~\ref{sec:other_operators}.

A convenient choice for dimension-6 operators is provided by the
``Higgs basis''~\cite{deFlorian:2016spz,Falkowski:2001958} in which the Higgs is assumed to be part of an ${\rm SU}(2)_L$ doublet and
operators connected to the LHC Higgs
searches are separated from the others that can be tested in observables not involving the Higgs.\footnote{For the relation between the independent couplings in the Higgs basis and the Wilson coefficients of other operator bases, see~\cite{Falkowski:2001958}.} The $10$ effective
operators we will focus on can be split into three classes:
the first one contains deformations of the Higgs couplings to the SM gauge bosons,
parametrized by
\begin{equation}\label{eq:H_VV}
\delta c_z\,, \  c_{zz}\,, \ c_{z\square}\,, \ \hat c_{z\gamma}\,, \ \hat c_{\gamma \gamma}\,, \ \hat c_{gg}\,,
\end{equation}
the second class is related to deformations of the fermion Yukawa's
\begin{equation}\label{eq:H_ff}
\delta y_{t}\,, \ \delta y_{b}\,,\ \delta y_{\tau}\,,
\end{equation}
and finally the last effect is a distortion of the Higgs trilinear self-coupling
\begin{equation}
\kappa_\lambda\,.
\end{equation}
The corresponding corrections to the Higgs interactions in the unitary gauge are given by
\begin{align}\nn
\mathcal{L} \supset & \, \frac{h}{v} \Bigg[ \delta c_w \frac{g^2 v^2}{2} W_{\mu}^+W^{-\mu} + \delta c_z \frac{(g^2 + g'^2) v^2}{4} Z_\mu Z^\mu \\\nn
&+c_{ww}\frac{g^2}{2}W_{\mu\nu}^+W^{-\mu\nu} + c_{w\square} g^2\left(W_{\mu}^-\partial_\nu W^{+\mu\nu} + \text{h.c.}\right) + \hat c_{\gamma\gamma}\frac{e^2}{4 \pi^2}A_{\mu\nu}A^{\mu\nu} \\\nn
& +c_{zz}\frac{g^2 + g'^2}{4} Z_{\mu\nu}Z^{\mu\nu}+ \hat c_{z\gamma}\frac{e \sqrt{g^2 + g'^2}}{2 \pi^2}Z_{\mu\nu}A^{\mu\nu}
+c_{z\square}g^2Z_\mu\partial_\nu Z^{\mu\nu}
+c_{\gamma \square}g g' Z_\mu\partial_\nu A^{\mu\nu}
\Bigg]\\\nn
&+ \frac{g_s^2}{48 \pi^2} \left(\hat c_{gg} \frac{h}{v} + \hat c_{gg}^{(2)} \frac{h^2}{2 v^2}\right) G_{\mu\nu} G^{\mu\nu}
-\sum_f \left[ m_f \left(\delta y_f \frac{h}{v} + \delta y_f^{(2)} \frac{h^2}{2v^2}\right) \bar{f}_{R}f_{L}+\text{h.c.}\right]\\
& - (\kappa_\lambda - 1) \lambda_3^{SM} v h^3\,, \label{eq:coupl_def}
\end{align}
where the parameters $\delta c_w$, $c_{ww}$, $c_{w\square}$, $c_{\gamma\square}$, $\hat c_{gg}^{(2)}$
and $\delta y_f^{(2)}$ are dependent quantities, defined as
\begin{alignat}{2}\nn
&\delta c_w & =\; & \delta c_z\,, \\\nn
& c_{ww} & =\; & c_{zz} + 2 \frac{g'^2}{\pi^2 (g^2 + g'^2)} \hat c_{z\gamma} + \frac{g'^4}{\pi^2  (g^2 + g'^2)^2} \hat c_{\gamma \gamma}\,,\\\nn
& c_{w\square} & =\; & \frac{1}{g^2 - g'^2}\Big[g^2 c_{z\square} + g'^2 c_{zz} - e^2 \frac{g'^2}{\pi^2  (g^2 + g'^2)} \hat c_{\gamma\gamma}
-(g^2-g'^2) \frac{g'^2}{\pi^2 (g^2 + g'^2)} \hat c_{z\gamma} \Big]\,,\\\nn
&c_{\gamma\square} & =\; & \frac{1}{g^2 - g'^2}\Big[2 g^2 c_{z\square} + \left(g^2 +g'^2\right)c_{zz}
- \frac{e^2}{\pi^2} \hat c_{\gamma\gamma}  - \frac{g^2-g'^2}{\pi^2}\hat c_{z\gamma} \Big]\,,\\\nn
& \hat c_{gg}^{(2)} & =\; & \hat c_{gg}\,,\\
& \delta y_f^{(2)} & =\; & 3 \delta y_f - \delta c_z\,.\label{eq:dependent_params}
\end{alignat}
In the above expressions we denoted by $g$, $g'$, $g_s$ the ${\rm SU}(2)_L$, ${\rm U}(1)_Y$ and ${\rm SU}(3)_c$ gauge couplings
respectively. The electric charge $e$ is defined by the expression $e = g g'/\sqrt{g^2 + g'^2}$.

Notice that in the Higgs basis the distortion of the
trilinear Higgs coupling is encoded in the parameter $\delta \lambda_3$ and denotes an additive shift in the
coupling, ${\cal L}_{\rm self} \supset - (\lambda_3^{SM} + \delta \lambda_3) v h^3$. In our notation $\kappa_\lambda$ denotes
instead a rescaling of the Higgs trilinear coupling, as specified in eq.~(\ref{eq:coupl_def}). We use this modified notation
in order to make contact with previous literature discussing the measurement of the Higgs self-coupling.

In eqs.~\eqref{eq:coupl_def} and~\eqref{eq:dependent_params} we also used a non-standard normalization
for the $\hat c_{gg}$, $\hat c_{\gamma\gamma}$
and $\hat c_{z\gamma}$ parameters. The contact Higgs coupling to gluons has been normalized to the
LO top loop prediction in the SM computed in the infinite $m_t$ limit, whereas we included an additional
factor $1/\pi^2$ in the couplings $\hat c_{\gamma\gamma}$ and $\hat c_{z\gamma}$.
The relation with the standard normalization of ref.~\cite{deFlorian:2016spz} is given by
\begin{equation}
c_{gg} = \frac{1}{12 \pi^2} \hat c_{gg} \simeq 0.00844 \hat c_{gg}\,,
\quad
c_{\gamma\gamma} = \frac{1}{\pi^2} \hat c_{\gamma\gamma} \simeq 0.101 \hat c_{\gamma\gamma}\,,
\quad
c_{z\gamma} = \frac{1}{\pi^2} \hat c_{z\gamma} \simeq 0.101 \hat c_{z\gamma}\,.
\end{equation}
With these normalizations values of order one for $\hat c_{gg}$, $\hat c_{\gamma\gamma}$ and $\hat c_{z\gamma}$ correspond
to BSM contributions of the same order of the SM gluon fusion amplitude and of the $H \rightarrow \gamma \gamma$
and $H \rightarrow Z \gamma$ partial widths.

\medskip

Since our analysis takes into account NLO corrections to the single-Higgs production and decay rates,
it is important to discuss the issue of renormalizability in our EFT setup.
In general, when we deform the SM Lagrangian with higher-dimensional operators, a careful renormalization procedure is needed
when computing effects beyond the LO. However, as discussed in ref.~\cite{Degrassi:2016wml},
if we are only interested in NLO effects induced by a modified Higgs trilinear self-coupling, no UV divergent contributions are
generated. This is a consequence of the fact that the Higgs trilinear coupling does not enter at LO
in single-Higgs observables but only starts to contribute at NLO. As far as the modified trilinear is concerned, our setup
essentially coincides with that of ref.~\cite{Degrassi:2016wml}, so we can carry over to our framework their results.
We report them in appendix~\ref{app:rates} for completeness.

Possible subtleties could instead arise considering the NLO contributions due to deformations of the single-Higgs couplings,
since these interactions already enter in the LO contributions. The deviations in single-Higgs couplings, however,
are already constrained to be relatively small, and will be tested in the future with a precision of the order of $10\%$ or below.
Their contributions at NLO can thus be safely neglected. For this reason we will include their effects only at LO, in which case no subtleties about renormalization arise.

\subsection{Additional operators contributing to Higgs observables}\label{sec:other_operators}

As we already mentioned, a larger set of dimension-6 operators can in principle affect Higgs observables.
We will list them in the following and discuss how they can be constrained through measurements not involving
the Higgs.
\begin{itemize}
\item {\it Vertex corrections.}
A first class of operators include the vertex corrections mediated by interactions of the form
\begin{equation}
{\cal O}_{vert} = (i H^\dagger \lra{D}_\mu H) (\overline f \gamma^\mu f)\,,
\qquad
{\cal O}^{(3)}_{vert} = (i H^\dagger \sigma^a \lra{D}_\mu H) (\overline f \gamma^\mu \sigma^a f)\,.
\end{equation}
They give rise at the same time to deformations of the couplings of the $Z$ and $W$ bosons with the fermions
and  to {\it hVf\,f} contact interactions. Both these effects can modify Higgs physics at tree level.
The gauge couplings deformations, for instance, affect the production cross section in vector boson fusion.
The {\it hVf\,f} vertices, instead, modify the cross section of $ZH$ and $WH$ production and the decay rates
in the $h \rightarrow VV^* \rightarrow 4f$ channels.

Under the assumption of flavor universality, all the vertex-correction operators can be constrained at the
$10^{-2} - 10^{-3}$ level~\cite{Pomarol:2013zra,Efrati:2015eaa,Bylund:2016phk}. Even in the high-luminosity LHC phase,
Higgs observables will have at least {\it few} $\%$ errors. Vertex corrections in flavor universal theories are thus too small
to be probed in Higgs physics and can be safely neglected.

If the assumption of flavor universality is relaxed, larger corrections to specific vertex operators
are allowed~\cite{Efrati:2015eaa}.\footnote{In this discussion we do not consider new-physics contributions to
the $W$ boson couplings with the right-handed fermions. Contributions induced by these couplings do not interfere
with the SM amplitudes and are thus typically too small to play any significant role.}
The gauge couplings involving leptons are still very well constrained
and below detection in Higgs physics. Sizable corrections can instead modify the quark couplings. In particular
the couplings involving the first generation quarks can deviate at the level of {\it few} $\%$ and Higgs measurements
at the high-luminosity LHC could be sensitive to them. The gauge couplings involving second generation quarks
or the bottom are still very well constrained. Finally the couplings involving the top quark are very poorly bounded.
In particular the coupling $Z \overline t_R t_R$ at present is practically unconstrained, while in the future it could
be tested with some accuracy in $t\overline tZ$ production.

\item {\it Dipole operators.} A second class of operators that can correct Higgs observables are dipole-like contact
interactions of the generic form
\begin{equation}
{\cal O}_{dip} = \overline f H  \sigma_{\mu\nu} T^a f F^{a\,\mu\nu}\,.
\end{equation}
These operators induce at the same time dipole interactions of the gauge bosons with the fermions and
vertices of the form {\it h\,$\partial$\!Vf\,f}.
The {\it h\,$\partial$\!Vf\,f} operators can modify Higgs decays into four fermions. However in this case the dipole contributions
do not interfere with the SM amplitudes since they have a different helicity structure. Moreover the experimental
bounds on dipole moments put strong constraints on the coefficients of the dipole operators, in particular for the
light generation fermions. For these reasons dipole operators can typically be neglected in Higgs physics~\cite{Falkowski:2015fla}.
A possible exception is the chromomagnetic operator involving the top quark, which can modify the $t\overline t H$ production
channel. Although in many BSM scenarios this operator is expected to be safely small, the current direct bounds from
the $t\overline t$ process are relatively weak~\cite{Franzosi:2015osa}, so that the top dipole operator could still
play a role in Higgs physics~\cite{Maltoni:2016yxb}.

\item {\it Four-fermion operators.} A third set of operators that can affect Higgs physics is given by four-fermion interactions.
Operators involving light generation fermions and the top quark can correct at tree-level the $t\overline tH$ production
channel. These effects are suppressed in several BSM scenarios
since they would be correlated to $4$-fermion interactions involving only light quarks, which are tightly constrained
by dijet searches. However the direct bounds on operators involving top quarks, which can be tested in $t\overline t$ production,
are not strong enough yet to forbid non-negligible effects in Higgs physics.\footnote{We thank E.~Vryonidou
for pointing this out to us.}

On the other hand, $4$-fermion operators involving only third generation fermions do not modify Higgs observables at tree-level,
but can induce loop corrections.
Obviously the loop factor gives a strong suppression for these effects. Nevertheless four-fermion operators involving
the top quark are poorly constrained at present, so that large coefficients are allowed, which could compensate the
loop suppression. For instance four-top operators can correct the gluon-fusion cross section, while
operators with top and bottom quarks can modify the Higgs branching ratio into a bottom pair.

Taking into account
the possible chirality structures, $12$ four-fermion operators involving only third generation quarks can be written.
A few constraints on some combination of them are available at present. The strongest one comes from the
measurement of the $Z \overline b_L b_L$ vertex, which receives loop corrections from four-fermion operators
involving the left-handed quark chirality~\cite{Elias-Miro:2013mua}. Additional constraints can be obtained from
bounds on the $\overline t t$ and $\overline t t \overline t t$ cross sections. For instance the current LHC
measurements put a bounds of order $1/(600\,{\rm GeV})^2$ on the coefficient of the
$(\overline t_R \gamma^\mu t_R)(\overline t_R \gamma_\mu t_R)$ operator~\cite{ATLAS:2016btu}.
A suppression of this size is enough to ensure that the loop corrections to Higgs physics are smaller than the
achievable precision.

Of course a fully model-independent analysis of the four-fermion operators should be done by considering all
operators simultaneously and not just one at a time (as done in the experimental analysis of ref.~\cite{ATLAS:2016btu}).
Such study is beyond the scope of this paper, so we will neglect the effects of four-fermion operators in our analysis.

\end{itemize}

A final comment is in order. In the above discussion we assumed that the BSM effects are parametrized
by dimension-6 operators in which the electroweak symmetry is linearly realized. This assumption allows to
relate the {\it hVf\,f} and {\it h\,$\partial$\!Vf\,f} operators to the vertex and dipole operators, so that these operators can be tested
in processes not involving the Higgs. If the electroweak symmetry is not linearly realized (or equivalently if the expansion in
Higgs powers is not valid) the interactions involving the Higgs become independent and can not be constrained any more
in non-Higgs physics. In such case a more complicated analysis, taking into account all the operators, must be performed.
We will give more details about the non-linear Lagrangian in the following subsection.

\subsection{Large Higgs self-interactions in a consistent EFT expansion}
\label{sec:EFT_validity}

An important issue to take into account when using the effective framework is the range of validity of the EFT approximation.
This is a delicate issue, crucially depending on the choice of power counting encoding the assumptions about the UV dynamics.
Here we only include a concise discussion with a few examples and refer the reader to the literature~\cite{Contino:2016jqw}
for possible subtleties.

As we will see in the following, the LHC measurements, especially in the high-luminosity phase, can probe inclusive single-Higgs
observables with a precision of the order or slightly below $10\%$. In the absence of new physics, possible BSM effects
will thus be constrained to be significantly smaller than the SM contributions. This translates into tight bounds on the
coefficients of the operators that correct the Higgs interactions with the gauge bosons (eq.~(\ref{eq:H_VV})) and with the
fermions (eq.~(\ref{eq:H_ff})). The leading effects due to these operators arise from the interference with the SM amplitude, while
quadratic terms are subleading. Corrections arising from dimension-8 operators lead to effects that are generically
of the same order of the square of the dimension-6 ones and are subleading as well.\footnote{There exist particular classes
of theories in which the size of effects coming from the dimension-8 operators is enhanced with respect to the square
of the dimension-6 ones. We will not consider these scenarios in our analysis. For a discussion of these effects see
for instance refs.~\cite{Azatov:2015oxa,Contino:2016jqw}.} This justifies our approximation of keeping only the leading EFT operators.

The discussion about the trilinear Higgs self-coupling is instead more subtle. As we will see in the following, the constraints
on $\kappa_\lambda$ we can obtain from the LHC data are quite loose. The Higgs trilinear coupling can only be tested at order
one, even at the end of the high-luminosity LHC program. Such large deviations in $\kappa_\lambda$, accompanied by small deviations in the
Higgs couplings to gauge fields and fermions, can only be obtained in very special BSM scenarios.
Indeed in generic new-physics models the deviations in all Higgs couplings are expected to be roughly of the same order.
For instance in models that follow the SILH power counting~\cite{Giudice:2007fh,Panico:2015jxa,Azatov:2015oxa} we expect
\begin{equation}
\delta c_z \sim v^2/f^2\,, \qquad \delta \kappa_\lambda \equiv \kappa_\lambda - 1 \sim v^2/f^2\,,
\end{equation}
where the $f$ parameter is related to the typical coupling $g_*$ and mass scale $m_*$ of the new dynamics
by $f \sim m_*/g_*$. In this class of models the deviations in the Higgs self-interactions are typically small, much below
the LHC sensitivity. A fit of the single-Higgs couplings, neglecting the trilinear Higgs modifications is thus fully justified
in these scenarios. At the same time the constraints achievable on $\kappa_\lambda$ at the LHC will hardly have
any impact in probing the parameter space of SILH theories.
The situation could however change at future high-energy machines, as for instance a $100$~TeV hadron collider,
which could test $\kappa_\lambda$ with a precision below $10\%$, implying non-trivial constraints
on models following the SILH power counting~\cite{Azatov:2015oxa,Contino:2016spe}.

Enhanced deviations only in Higgs self-couplings are possible in other classes of models. Interesting possibilities
are provided for instance  (i) by scenarios in which the Higgs is a generic bound state of a strongly coupled dynamics (i.e.~not a Goldstone
boson) (see discussion in ref.~\cite{Azatov:2015oxa}), (ii) by bosonic technicolor scenarios and (iii) by Higgs-portal models.
In all these cases large deviations in the Higgs self-couplings can be present
and accompanied by small corrections in single Higgs interactions. As an explicit example, we will analyze
the Higgs portal scenarios later on.

It is important to stress that, in the presence of large corrections to Higgs self-interactions, the EFT expansion
in Higgs field insertions may break down. In this case the expansion in derivatives can still be valid, since it is controlled by the
expansion parameter $E/\Lambda$, but we can not neglect operators with arbitrary powers of the Higgs field. The effective
parametrization can still be used in such situation provided that we interpret the effective operators as a ``resummation'' of
the effects coming from operators with arbitrary Higgs insertions. This is equivalent to a ``non-linear'' effective
parametrization in which the Higgs is not assumed to be part of an ${\rm SU}(2)_L$ doublet, but is instead treated as a full singlet (see ref.~\cite{deFlorian:2016spz} for a brief account on non-linear EFT and for a list of further references).
The only caveat with this parametrization is the fact that interactions with multiple Higgs fields are not connected any more
to the single-Higgs couplings. In this case a different global fit should be performed,
in which $c_{gg}^{(2)}$ and $\delta y_f^{(2)}$ are treated as independent parameters. Notice also that 
the {\it hVf\,f} and {\it h\,$\partial$\!Vf\,f} operators should a priori be included in the analysis, as we discussed in sec.~\ref{sec:other_operators} and EW precision data and Higgs data cannot be analyzed separately any longer.

\medskip

To clarify the issues discussed above, we now analyze an explicit class of models, the Higgs portal scenarios.
As a concrete example, we assume that a new scalar singlet $\varphi$, neutral under the SM gauge group, is
described by the Lagrangian\footnote{The power counting we derive in the following applies also to more general
Higgs portal models. In particular it is valid for scenarios characterized by a single coupling $g_*$ and a single mass scale $m_*$
in which the Higgs is coupled to the new dynamics through interactions of the type $\theta H^\dagger H {\cal O}$,
where $\cal O$ is a generic new-physics operator. Note that a different power counting can arise for portal scenarios in which
the new-physics sector is charged under the SM (see ref.~\cite{deBlas:2014mba} for a classification of possible scenarios).}
\begin{equation}
{\cal L} \supset \theta g_* m_* H^\dagger H \varphi - \frac{m_*^4}{g_*^2}\, V\!\left({g_* \varphi}/{m_*}\right)\,,
\end{equation}
where the dimensionless parameter $\theta$ measures the mixing between the Higgs sector and the neutral sector, and $V$ is a generic potential.
In the EFT description obtained after integrating out $\varphi$
the derivative expansion is valid if $E/m_* \ll 1$, while the expansion in Higgs-field
insertions is valid when
\begin{equation}
\varepsilon \equiv \frac{\theta g_*^2 v^2}{m_*^2} \ll 1\,.
\end{equation}
Note that $\theta$ and $\varepsilon$ are  truly dimensionless quantities in mass and coupling dimensions.
The corrections to the Higgs couplings with gauge fields come indirectly from operators of the type
$\partial_\mu (H^\dagger H)\partial^\mu (H^\dagger H)$ and can be estimated as
\begin{equation}
\delta c_z \sim \theta^2 g_*^2 \frac{v^2}{m_*^2}\,.
\end{equation}
The corrections to the Higgs trilinear coupling are instead given by
\begin{equation}\label{eq:dev_kappa}
\delta \kappa_\lambda \sim \theta^3 g_*^4 \frac{1}{\lambda_3^{SM}} \frac{v^2}{m_*^2}\,.
\end{equation}
First of all, we can notice that $\delta \kappa_\lambda \sim \theta g_*^2/\lambda_3^{SM} \delta c_z$, thus a large hierarchy
between the corrections to linear Higgs couplings and the deviation in the self-interactions requires
sizable values of the Higgs portal coupling $\theta$ (and/or large values of the new-sector coupling $g_*$).

When the corrections to the Higgs potential become large, some amount of tuning is typically needed to fix the correct
properties of the Higgs potential.
Notice that Higgs-portal scenarios do not typically provide a solution to the hierarchy problem. Thus they will in general
suffer from some amount of tuning in the Higgs mass term, exactly as generic extensions of the SM. On top of this
some additional tuning in the Higgs quartic coupling can also be present.
In the following we will refer only to this additional tuning, which we denote by $\Delta$.
We can estimate $\Delta$ by noticing that the quartic coupling needs to be fixed with a precision of the order
of $\lambda_3^{SM}$. By comparing the new-physics corrections to the quartic coupling with the SM value we get 
\begin{equation}
\Delta \sim \frac{\theta^2 g_*^2}{\lambda_3^{SM}}\,.
\end{equation}
We can easily relate $\delta \kappa_\lambda$ given in eq.~(\ref{eq:dev_kappa}) to the amount of tuning $\Delta$ as
\begin{equation}
\delta \kappa_\lambda \sim \varepsilon \Delta.
\end{equation}
This relation has an interesting consequence. If we require the expansion in Higgs insertions to be valid ($\varepsilon \lesssim 1$)
and the model not to suffer additional tuning ($\Delta \lesssim 1$), we get that the corrections to the Higgs trilinear coupling can
be at most of order one ($\delta \kappa_\lambda \lesssim 1$). Larger corrections can however be obtained if at least one of the two
conditions $\varepsilon \lesssim 1$ and $\Delta \lesssim 1$ is violated.

As we already mentioned, if the expansion in Higgs insertions is not valid ($\varepsilon > 1$),
large deviations in the Higgs couplings are possible.
In particular single- and multiple-Higgs couplings are not related any more and a non-linear effective parametrization
must be used. In this scenario, however, large corrections to the linear Higgs couplings to the SM fields are expected, so that
significant tuning is required to pass the precision constraints from single-Higgs processes.

A second scenario, in which $\varepsilon \lesssim 1$ while some tuning is present in the Higgs potential ($\Delta > 1$),
can instead naturally lead to small deviations in the linear Higgs couplings. For instance by taking
$\theta \simeq 1$, $g_* \simeq 3$ and $m_* \simeq 2.5$~TeV we get
\begin{equation}
\varepsilon \simeq 0.1\,, \quad 1/\Delta \simeq 1.5\%\,,
\quad \delta c_z \simeq 0.1\,, \quad \delta \kappa_\lambda \simeq 6\,.
\end{equation}

Since we are going to consider sizable deviations in the trilinear Higgs coupling, it is important to understand whether
such corrections are compatible with a high-enough cut-off of the effective description.
If large corrections are present in the Higgs self-interactions, scattering processes involving
longitudinally polarized vector bosons and Higgses, in particular $V_L V_L \rightarrow V_L V_L h^n$, lose perturbative unitarity
at relatively low energy scales.  The upper bound for the cut-off of the EFT description can be estimated as~\cite{Falkowski,FalkowskiRattazzi}
\begin{equation}
\Lambda \lesssim \frac{4 \pi v}{\sqrt{|\kappa_\lambda - 1|}} \sqrt{\frac{32 \pi}{15}} \frac{v}{m_h}\,.
\end{equation}
This bound is not very stringent: for $|\kappa_\lambda| \lesssim 10$ one gets $\Lambda \lesssim 5$~TeV.
For values of $\kappa_\lambda$ within the expected high-luminosity LHC bounds, perturbativity loss is thus well above
the energy range directly testable at the LHC.

As a last point, we comment on the issue of the stability of the Higgs vacuum. As pointed out in ref.~\cite{Degrassi:2016wml}, if the only deformation
of the Higgs potential is due to the $(H^\dagger H)^3$ operator, the usual vacuum is not a global minimum for
$\kappa_\lambda \gtrsim 3$. In this case the vacuum becomes metastable, although it could still have a long enough lifetime.
Additional deformations from higher-dimensional operators can remove the metastability bound, even for large values of $\kappa_\lambda$. A lower bound $\kappa_\lambda > 1$ can also be extracted if we naively require the Higgs potential
to be bounded from below for arbitrary values of the Higgs VEV $\langle h \rangle$,
i.e.~if we require the coefficient of the $(H^\dagger H)^3$ operator to be positive.
This constraint, however, is typically too restrictive. Our estimate of the effective potential, in fact, is only valid
for relatively small values of the Higgs VEV, which satisfy
$\varepsilon = \theta g_*^2 \langle h \rangle^2/m_*^2 \lesssim 1$. For large values of $\langle h \rangle$
the expansion in the Higgs field breaks down and the estimate of the potential obtained by including only dimension-$6$ operators
is not reliable any more and the whole tower of higher-dimensional operators should be considered.
In this case large negative corrections to the Higgs trilinear coupling could be compatible with a stable vacuum.
Examples of such scenarios are the composite Higgs models in which the Higgs field is identified with a Goldstone boson.
In these models the Higgs potential is periodic and a negative coefficient for the effective $(H^\dagger H)^3$ operator
does not generate a runaway behavior of the potential.

\section{Fit from inclusive single-Higgs measurements}\label{sec:inclusive}

As we mentioned in the introduction, single-Higgs production measurements can be sensitive to large variations
of the Higgs trilinear self-coupling. These effects arise at loop level and
can be used to extract some constraints on the $\kappa_\lambda$ parameter. Under the assumption that only
the trilinear Higgs coupling is modified, $\kappa_\lambda$ can be constrained to the range
$\kappa_\lambda \in [-0.7, 4.2]$ at the $1\sigma$ level and
$\kappa_\lambda \in [-2.0, 6.8]$ at $2\sigma$~\cite{Degrassi:2016wml} at the end of the high luminosity phase of the LHC. This result was obtained by assuming that the
experimental uncertainties are given by the `Scenario 2' estimates of CMS~\cite{CMS:2013xfa,Peskin:2013xra},
in which the theory uncertainties are halved with respect to the 8\,TeV LHC run and the other systematic uncertainties
are scaled as the statistical errors. The actual precision achievable
in the high-luminosity LHC phase could be worse than this estimate, leading to a slightly smaller sensitivity
on $\kappa_\lambda$. Nevertheless the result shows that single Higgs production could be competitive with
other measurements, for instance double-Higgs production, in the determination of the Higgs self
coupling.

A similar analysis, focusing only on the gluon fusion cross section and on the $H \rightarrow \gamma\gamma$
branching ratio, was presented in ref.~\cite{Gorbahn:2016uoy}. With this procedure a bound $\kappa_\lambda \in [-7.0, 6.1]$
at the $2\sigma$ level was derived, whose overall size is in rough agreement with the result of ref.~\cite{Degrassi:2016wml}.

In section~\ref{sec:EFT_validity} we saw that large corrections to the Higgs self-couplings are seldom generated alone
and are typically accompanied by deviations in the other Higgs interactions. In scenarios that predict
${\cal O}(1)$ corrections to $\kappa_\lambda$, single Higgs couplings, such as Yukawa interactions or couplings with the
gauge bosons, usually receive corrections of the order of $5 - 10 \%$. Since these corrections modify single-Higgs
processes at tree level, their effects are comparable with the ones induced at loop level by a modification of the
Higgs self-coupling. In these scenarios, a reliable determination of $\kappa_\lambda$ thus requires a global fit, in which
also the single-Higgs coupling distortions are properly included.

In this section we will perform such a fit, taking into account deformations of the SM encoded by the $10$ effective
operators introduced in section~\ref{sec:eft_parametrization} (see eq.~(\ref{eq:coupl_def})). As we will see, when all the
effective operators are turned on simultaneously, some cancellations are possible, resulting in an unconstrained
combination of the effective operators. This flat direction can not be resolved by taking into account only inclusive
single-Higgs production measurements. Additional observables are thus needed to disentangle deviations in the
Higgs self-coupling from distortions of single-Higgs interactions. We will discuss various possibilities along this line
in sections~\ref{sec:double_Higgs} and~\ref{sec:differential}.

Before performing the actual fit, it is also important to mention that large deviations in $\kappa_\lambda$ could
in principle also have an impact on the determination of single-Higgs couplings.
We will discuss this point in section~\ref{sec:effects_on_single}.

\subsection{Single-Higgs rates and single-Higgs couplings}\label{sec:single_Higgs_fit}

As a preliminary step in our analysis, we focus on single Higgs couplings neglecting the effects of $\kappa_\lambda$
and we perform a global fit exploiting single-Higgs processes.

Measurements of the production and decay rates of the Higgs boson are usually reported in terms of signal strengths,
i.e.~the ratio of the measured rates with respect to the SM predictions. The total signal strength, $\mu^f_i$,
for a given production mode $i$ and decay channel $h \rightarrow f$, is thus given by
\be
\mu_i^f = \mu_i \times \mu^f = \frac{\sigma_i}{(\sigma_i)_{\text{SM}}} \times \frac{\text{BR}[f]}{(\text{BR}[f])_\text{SM}}\,.
\ee
Obviously the production and decay signal strengths can not be separately measured and only
their products are directly accessible.

Single Higgs production can be extracted with good accuracy at the LHC in five main modes: gluon fusion (ggF),
vector boson fusion (VBF), associated production with a $W$ or a $Z$ ($WH$, $ZH$), and associated production
with a top quark pair ($t\overline tH$). Moreover the main Higgs decay channels are into $ZZ$, $WW$,
$\gamma\gamma$, $\tau^+\tau^-$ and $b\bar{b}$.\footnote{For simplicity we neglect the $\mu^+\mu^-$ and $c\bar{c}$
decay modes and assume that no invisible decay channels are present.}
A large subset of all the combinations of these production and decay modes can be
extracted at the high-luminosity LHC with a precision better than $10 - 20\%$. It is thus possible to linearly expand
the signal strengths as
\begin{equation}\label{eq:linear_mu}
\mu^f_i \simeq 1 + \delta \mu_i + \delta \mu^f\,,
\end{equation}
since quadratic terms are negligible.

As can be seen from eq.~(\ref{eq:linear_mu}), a rescaling of the production rates $\mu_i \rightarrow \mu_i + \delta$
can be exactly compensated by a rescaling of the branching ratios $\mu^f \rightarrow \mu^f - \delta$. For this reason,
out of the $10$ quantities describing the production and decay of an on-shell particle ($5$ productions and $5$ decays), only $9$ independent constraints can be
derived experimentally, which are enough to determine the set  of single-Higgs couplings $(\delta c_z, c_{zz}, c_{z\square}, \hat c_{z\gamma},
\hat c_{\gamma\gamma}, \hat c_{gg}, \delta y_t, \delta y_b, \delta y_\tau)$.

In our numerical analysis
we estimate the theory and experimental systematic uncertainties by following the ATLAS projections presented in
ref.~\cite{ATLAS:projections}. The full list of uncertainties is given in table~\ref{tab:errors}.
Notice that, with respect to the ATLAS analysis we introduced a few updates.
We reduced the theory uncertainty in the gluon fusion production cross section to take into account the recent improvement
in the theory predictions~\cite{Anastasiou:2016cez,deFlorian:2016spz}.
In addition, we updated the entries corresponding to the VBF production mode with $ZZ$ final state using the more
recent estimates presented in ref.~\cite{ATL-PHYS-PUB-2016-008}.
To estimate the separate uncertainties in the $WH$ and the $ZH$ production modes with $ZZ$ final state,
which are considered together in ref.~\cite{ATLAS:projections},
we divided the experimental uncertainty for $VH$ by the square root of the corresponding event
fractions.\footnote{In this way, we get that the ratio of uncertainties between the $WH$ and $ZH$ channels with $ZZ$ final state is in good agreement with a previous estimate by ATLAS~\cite{ATL-PHYS-PUB-2013-014}.}

Our projections are also in fair agreement with the `Scenario 1' in the CMS extrapolations~\cite{CMS:2013xfa},
in which the systematic uncertainties are assumed to be the same as in the 8\,TeV LHC run. Notice that our choice is more
conservative than the one made in ref.~\cite{Degrassi:2016wml}, and should be interpreted
as a `pessimistic' scenario. We will comment in section~\ref{sec:change_uncertainties} on how the numerical results change as a function of the systematic
uncertainties.

\begin{table*}
	\begin{center}
		\begin{tabular}{lc|ccc}
			\multicolumn{2}{c|}{Process}	 & Combination & Theory & Experimental \\ \hline \hline
			\multirow{5}{*}{${H}\to \gamma \gamma$}	&$\text{ggF}$  & 0.07 & 0.05 & 0.05 \\
			&$\text{VBF}$ & 0.22 & 0.16 & 0.15  \\
			&${t\overline tH}$& 0.17 & 0.12 & 0.12 \\
			&${WH}$& 0.19 & 0.08 & 0.17 \\
			&${ZH}$& 0.28 & 0.07 & 0.27 \\ \hline
			\multirow{5}{*}{${H}\to {ZZ}$}   		&$\text{ggF}$& 0.06 & 0.05 & 0.04 \\
			&$\text{VBF}$& 0.17 & 0.10 & 0.14 \\
			&${t\overline tH}$& 0.20 & 0.12 & 0.16 \\
			&${WH}$& 0.16 & 0.06 & 0.15 \\
			&${ZH}$& 0.21 & 0.08 & 0.20 \\ \hline
			\multirow{2}{*}{${H}\to {WW}$} 		&$\text{ggF}$& 0.07 & 0.05 & 0.05 \\
			&$\text{VBF}$& 0.15 & 0.12 & 0.09 \\\hline
			${H}\to {Z\gamma}$					&incl.		 & 0.30 & 0.13 & 0.27 		\\\hline
			\multirow{2}{*}{${H}\to {b\bar b}$}		&${WH}$& 0.37 & 0.09 & 0.36 \\
			&${ZH}$& 0.14 & 0.05 & 0.13 \\\hline
			${H}\to \tau^+ \tau^- $						&$\text{VBF}$ & 0.19 & 0.12 & 0.15
		\end{tabular}
		\caption{Estimated relative uncertainties on the determination of single-Higgs production channels at the high-luminosity LHC
		($14$~TeV center of mass energy, $3/{\rm ab}$ integrated luminosity and pile-up $140$~events/bunch-crossing). The theory, experimental (systematic
		plus statistic) and combined uncertainties are listed in the `Theory', `Experimental' and `Combination' columns
		respectively. All the estimates are derived from refs.~\cite{ATLAS:projections,ATL-PHYS-PUB-2016-008} and~\cite{Anastasiou:2016cez,deFlorian:2016spz}.}\label{tab:errors}
	\end{center}
\end{table*}

To extract the fit we assume that the central values of the measured signal strengths are equal to the SM predictions, i.e.~$\mu_i^f = 1$,
and we perform a simple statistical analysis by constructing the $\chi^2$ function
\be
\chi^2 =\sum_{i,f}\frac{(\mu_i^f - 1)^2}{(\sigma_i^f)^2}\,,
\ee
where $\sigma_i^f$ are the errors associated to each channel.

If we consider only small deviations in the single-Higgs couplings, we can linearly expand the signal strengths in terms
of the $9$ fit parameters (the numerical expressions are given in appendix~\ref{app:rates}).
In this way the $\chi^2$ function becomes quadratic in the parameters and we end up
in a Gaussian limit. The $1\sigma$ intervals and the full correlation matrix (with large correlations enlightened in boldface)  for the parameters are given by (by construction the
best fit coincides with the SM point, where all the coefficients vanish)
\be
\left(
\begin{array}{c}
\hat{c}_{gg} \\ \delta c_z \\ c_{zz} \\ c_{z\square} \\ \hat{c}_{z\gamma} \\ \hat{c}_{\gamma\gamma}\\ \delta y_t \\ \delta y_b \\ \delta y_\tau 
\end{array}
\right)
= \pm
\left(
\begin{array}{cc}
0.07 & (0.02)\\
0.07 & (0.01) \\
 0.64 & (0.02)\\
 0.24 & (0.01)\\
 4.94 & (0.65) \\
 0.08 & (0.02)\\ 
 0.09 & (0.02)\\
 0.14 & (0.03) \\
 0.17 & (0.09)
\end{array}
\right)
\quad\quad
\begingroup\makeatletter\def\f@size{7}\check@mathfonts
\left[
\begin{array}{ccccccccc}
 1 & -0.01 & -0.02 & 0.03 & 0.08 & 0.01 & \mathbf{-0.71} & 0.03 & 0.01 \\
 & 1 & -0.45 & 0.36 & -0.61 & -0.33 & 0.18 & \mathbf{0.89} & 0.53 \\
 &  & 1 & \mathbf{-0.99} & 0.69 & 0.11 & 0.38 & -0.47 & \mathbf{-0.74} \\
 &  &  & 1 & -0.58 & -0.23 & -0.42 & 0.42 & \mathbf{0.71} \\
 & & & & 1 & -0.58 & 0.09 & -0.46 & -0.63 \\
 & & & & & 1 & 0.14 & 0.04 & 0.04 \\
 & & & & & & 1 & 0.25 & -0.08 \\
& & & & & & & 1 & 0.57 \\
& & & & & & & & 1
\end{array}
\right]\,.
\endgroup
\label{eq:1sigmaunc}
\ee
The numbers listed in parentheses  correspond to the $1\sigma$ uncertainties
obtained by considering only one parameter at a time, i.e.~by setting to zero the coefficients of all the other effective operators.

The comparison between the global fit and the fit to individual operators shows that some bounds can significantly vary with the
two procedures. The most striking case, as noticed already in ref.~\cite{Falkowski:2015fla}, involves the $c_{zz}$
and $c_{z\square}$ coefficients, whose fit shows a high degree of correlation. As a consequence, the constraints obtained in the
global fit are more than one order of magnitude weaker than the individual fit ones. This high degeneracy can be lifted
by including in the fit constraints coming from EW observables. Indeed, as we will discuss later on, a combination of
the $c_{zz}$ and $c_{z\square}$ operators also modifies
the triple gauge couplings, generating an interesting interplay between Higgs physics and vector boson pair production.

Another element of particular interest in the correlation matrix is the $\hat c_{gg}$ -- $\delta y_t$ entry. The cleanest observable
constraining these couplings is the gluon fusion cross section, which however can only test a combination of the two parameters.
In order to disentangle them one needs to consider the $t\overline tH$ production mode.
This process, however, has a limited precision at the LHC, explaining the large correlation between $\hat c_{gg}$ and $\delta y_t$
and the weaker bounds in the global fit. Other ways to gain information about the top Yukawa coupling are to rely on an exclusive analysis
of gluon fusion with an extra hard jet~\cite{Grojean:2013nya} or to consider the effects of off-shell Higgs
production~\cite{Azatov:2014jga,Azatov:2016xik}

High correlations are also present between the bottom Yukawa parameter $\delta y_b$ and all the other parameters
except $\hat c_{gg}$ and $\delta y_t$. The origin of the correlations can be traced back to the fact that the main impact
of a modified bottom Yukawa is a rescaling of the Higgs branching ratios. Since the $b\overline b$ decay channel can only be
tested with limited accuracy, the main constraints on $\delta y_b$ come exploiting the gluon fusion channel
with the Higgs decaying into $\gamma \gamma$, $ZZ$, $WW$ and $\tau \tau$. A variation of the bottom Yukawa leaves the gluon
fusion cross section nearly unchanged, thus to recover the SM predictions one needs to compensate the variations in the
branching ratios induced by $\delta y_b$ with contributions from the $\delta c_z$, $c_{zz}$, $c_{z\square}$, $\hat c_{z\gamma}$,
$\hat c_{\gamma\gamma}$ and $\delta y_\tau$. This feature gives rise to the large correlations between $\delta y_b$ and these
parameters.

\medskip

The presence of sizable correlations among various parameters significantly limits the robustness of the results shown in
eq.~(\ref{eq:1sigmaunc}). In particular the Gaussian approximation we used to derive the bounds is not fully justified.
We checked that, by using the full expressions for the signal rates the $1\sigma$ limits are significantly modified.
The largest effects are found in the $c_{zz}$ and $c_{z\square}$ bounds, which change at order one.
Such large sensitivity to the quadratic (and higher-order) terms in the fit also signals that corrections coming from
higher-dimensional effective operators could also affect the fit in a non-negligible way. To solve this problem we need to
lift the approximate flat directions related to the large entries in the correlation matrix.  One way to achieve this goal is to
include in the fit additional observables that can provide independent constraints on the Higgs couplings.
We will list in the following a few possibilities.

{\bf Di-boson data.} A first set of observables that has an interplay with Higgs physics is given by
the EW boson trilinear gauge couplings (TGC's). In the Higgs basis the deviations of two TGC's are correlated to the single-Higgs
couplings modifications.
Measurements of the $W W Z$ and $W W \gamma$ interactions can be converted into constraints on two linear combinations of the  $\hat c_{\gamma\gamma}$, $\hat c_{z\gamma}$, $c_{zz}$ and $c_{z\square}$ parameters (see the explicit expressions in appendix~\ref{app:trilinear}),
which can be used to remove the correlation between $c_{zz}$ and $c_{z\square}$. At present the $WWZ$ and $WW\gamma$
couplings are tested with an accuracy of order $\sim 5\%$~\cite{Butter:2016cvz,Falkowski:2016cxu}.
For our numerical
analyses we will assume a precision of order $1\%$ at the end of the high-luminosity LHC phase.

{\bf Rare Higgs decays.} Another set of observables related to the Higgs couplings is obtained by considering additional, more rare Higgs decays.
The inclusion of the $h\to Z\gamma$ decay, which is expected to be measured with
$\sim 30\%$ accuracy at the high-luminosity LHC~\cite{ATL-PHYS-PUB-2014-006}, can be used to constrain the $\hat c_{z\gamma}$ parameter.
The $h\to \mu^+\mu^-$ decay, on the other hand, has a limited impact on the fit, since it depends on an additional
parameter, the deviation in the muon Yukawa $\delta y_\mu$.
In the flavor universal case, however, the muon and tau Yukawa receive equal new-physics contributions, $\delta y_\mu = \delta y_\tau$,
and the determination of $\delta y_\mu$ can be used to improve the fit on $\delta y_\tau$. The improvement is anyhow
limited, since the precision achievable in the measurement of the
$h \to \mu^+\mu^-$ decay is comparable with the one achievable directly on the $\tau$ Yukawa.
Apart from the impact on $\delta y_\tau$, the influence of the $h\to \mu^+\mu^-$ channel
on the fit of the remaining single Higgs couplings is negligible.

The above constraints, in particular the ones coming from TGC's and $h \rightarrow Z\gamma$, significantly help in
improving the fit on single Higgs couplings and lowering the correlations. The $1\sigma$ fit intervals on the EFT parameters
and the correlation matrix are modified as
\be
\left(
\begin{array}{c}
\hat{c}_{gg} \\ \delta c_z \\ c_{zz} \\ c_{z\square} \\ \hat{c}_{z\gamma} \\ \hat{c}_{\gamma\gamma}\\ \delta y_t \\ \delta y_b \\ \delta y_\tau 
\end{array}
\right)
= \pm
\left(
\begin{array}{cc}
0.07 & (0.02)\\
0.05 & (0.01) \\
 0.05 & (0.02)\\
 0.02 & (0.01)\\
 0.09 & (0.09) \\
 0.03 & (0.02)\\ 
 0.08 & (0.02)\\
 0.12 & (0.03) \\
 0.11 & (0.09)
\end{array}
\right)
\quad\quad
\begingroup\makeatletter\def\f@size{7}\check@mathfonts
\left[
\begin{array}{ccccccccc}
 1 & 0.04 & -0.01 & -0.01 & 0.04 & 0.31 & \mathbf{-0.76} & 0.05 & 0.02 \\
 & 1 & -0.07 & -0.26 & 0.01 & 0.01 & 0.36 & \mathbf{0.88} & 0.27 \\
 & & 1 & \mathbf{-0.87} & 0.13 & 0.20 & 0.03 & -0.07 & -0.06 \\
 & & & 1 & -0.09 & -0.09 & -0.09 & -0.17 & 0.08 \\ 
 & & & & 1 & 0.05 & -0.02 & -0.02 & -0.03 \\
 & & & & & 1 & -0.32 & -0.19 & -0.12 \\
 & & & & & & 1 & 0.50 & 0.28 \\
 & & & & & & & 1 & 0.36 \\
 & & & & & & & & 1 \\
\end{array}
\right]\,.
\endgroup
\label{eq:1sigmaunc_tgcconstr}
\ee
These results have been obtained by linearizing the signal strengths.
We however checked that, by using the full expressions for the $\mu_i^f$, the results in
eq.~(\ref{eq:1sigmaunc_tgcconstr}) remain basically unchanged.
The additional constraints coming from the TGC's and $h \rightarrow Z\gamma$ measurements thus effectively
resolve the approximate flat directions making our linearized EFT fit fully consistent and robust.

{\bf Higgs width.} Finally one could also consider the constraint on the Higgs total width, which could be extracted by comparing
off-shell and on-shell Higgs
measurements~\cite{Kauer:2012hd,Caola:2013yja,Campbell:2013una,Khachatryan:2014iha,Aad:2015xua}.\footnote{See also refs.~\cite{Englert:2014aca,Cacciapaglia:2014rla,Azatov:2014jga}
for possible issues related to the EFT interpretation of these measurements.} ATLAS estimated that a precision of $40\%$ could be reached at the end of the high-luminosity LHC~\cite{ATL-PHYS-PUB-2015-024}. If we include this piece of information in the fit, we find that also this constraint has a negligible impact on the flat directions. To assess whether an improvement on such projections could have an effect on the global fit, we repeated our analysis varying the estimated precision on the width. As expected, the most sensitive coefficients are $\delta y_b$ and $\delta c_z$. In order to affect their $1\sigma$ fit intervals, one needs a precision on the width of at least $20\%$. In particular, we find that if we assume a precision of $40\%$, $20\%$, and $10\%$, the $1\sigma$ bound on $\delta y_b$ of eq.~\eqref{eq:1sigmaunc_tgcconstr} shrinks to $0.11$, $0.09$, and $0.06$, while the one on $\delta c_z$ is reduced respectively to $0.05$, $0.04$ and $0.03$.

\medskip

To conclude the discussion about single-Higgs couplings, it is useful to report on what happens if we relax the assumption
of small deviations in the Higgs interactions. In this case the linear expansion in the signal strengths is no longer
appropriate and the full expressions must be retained. Additional minima
are then present in the fit. Trivial ones are obtained by reversing the sign of the tau ($\delta y_\tau \simeq - 2$)
or bottom ($\delta y_b \simeq - 2$) Yukawas, which leave the production cross sections and decay branching ratios
unchanged.\footnote{In the case of a `wrong-sign' botton Yukawa with an unchanged top Yukawa a small contribution
from $\hat c_{gg}$ is needed to compensate for the small change in the gluon fusion cross section.}
Other minima are obtained by choosing $\hat c_{gg}$
in such a way that its contribution to the gluon fusion amplitude is minus twice the SM one ($\hat c_{gg} \simeq - 2$)
or by choosing $\hat c_{\gamma\gamma}$ so that it reverses the amplitude for Higgs decay into
a photon pair ($\hat c_{\gamma\gamma} \simeq 1.6$).
Less trivial minima are instead obtained by reversing the top Yukawa coupling ($\delta y_t \simeq -2$), with either $\hat{c}_{gg}\simeq 0$ or $\hat{c}_{gg}\simeq 2$.
In this case the interference between the $W$ and top contributions to the branching ratio $h \rightarrow \gamma\gamma$
changes sign and must be compensated by a contribution from $\hat c_{\gamma\gamma}$
($\hat c_{\gamma\gamma} \simeq 2.1$ or $\hat c_{\gamma\gamma} \simeq 0.46$). An additional possibility is
to reverse the sign of the associated production channels amplitude ($\delta c_z \simeq -2$),
in which case the change in the $h \rightarrow \gamma\gamma$ amplitude can be compensated
by $\hat c_{\gamma\gamma} \simeq -0.45$ or $\hat c_{\gamma\gamma} \simeq -2.1$.
Finally by reversing both the sign of both the top Yukawa and of the associated production channels amplitude,
one finds two additional minima with $\hat c_{\gamma\gamma} \simeq -1.6$ or $\hat c_{\gamma\gamma} \simeq 0.01$.

Some of these additional minima can be probed by considering other observables. The sign of the top Yukawa can
be extracted from the measurement of $h$ + top production, as shown in refs.~\cite{Biswas:2012bd, Farina:2012xp,Demartin:2015uha}.
Large contributions to $\hat c_{gg}$ can instead be probed in double-Higgs production, which can be
used to exclude the $\hat c_{gg} \simeq -2$ minimum~\cite{Azatov:2015oxa}.
The sign of the bottom Yukawa can instead be tested by considering the transverse momentum distributions in Higgs production
with an extra jet~\cite{Bishara:2016jga} (see also ref.~\cite{Bonner:2016sdg}).\footnote{An additional Higgs associated
production channel, namely $H\gamma$, could be used to test large deviations in
$\hat c_{\gamma\gamma}$~\cite{Khanpour:2017inb}.}
We are instead not aware of any process which could be sensitive to the sign of the tau Yukawa.

In our analysis we also assumed that the sign of the $hWW$ and $hZZ$ couplings are the same (fixing them to be positive for
definiteness). Such assumption is well motivated theoretically, since a sign difference would imply large contributions to
custodial breaking operators. From the experimental point of view, however, testing the sign of the $hWW$ and $hZZ$ couplings
explicitly is very difficult at the LHC. It could be possible at future lepton colliders, which could be sensitive to the relative sign
of the two couplings in $ZH$ and $ZHH$ production~\cite{Farina:2012ea}.

\subsection{Global fit including Higgs self-coupling}\label{sec:global_fit_kappa}

We can now discuss how the above picture changes when we introduce in the fit the additional parameter $\kappa_\lambda$
controlling the Higgs self-coupling deformations.
As we saw in the previous subsection, the measurement of $5$ production
and $5$ Higgs decay channels allows us to extract $9$ independent constraints on the coefficients of the EFT Lagrangian.
By introducing $\kappa_\lambda$ in our fit, we reach a total of $10$ independent parameters, thus we expect one
linear combination to remain unconstrained in the fit. This is indeed what happens. The global fit
has an exact flat direction along which the $\chi^2$ vanishes.

\begin{figure}
\centering
\includegraphics[width=1\textwidth]{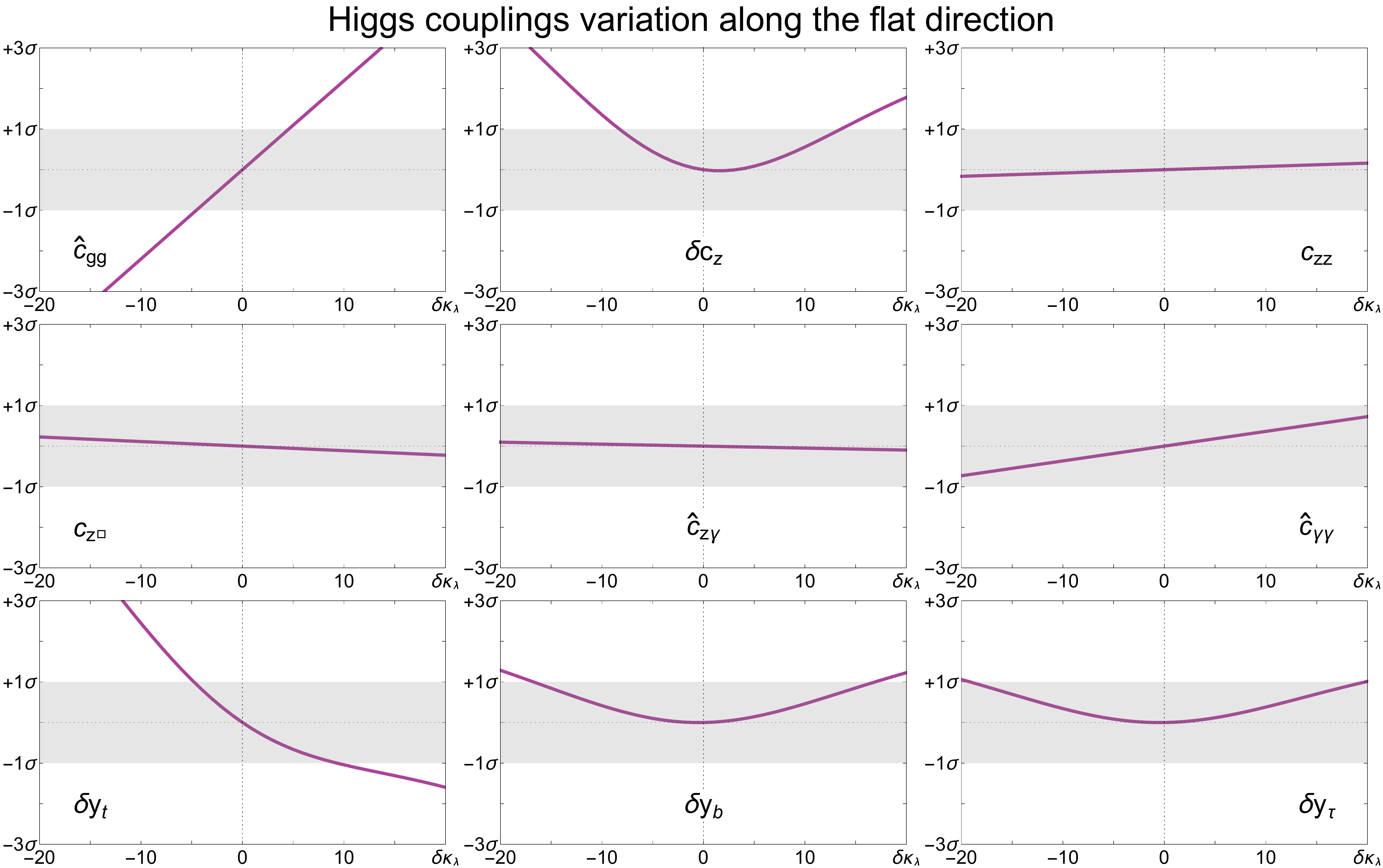}
\caption{Variation of the Higgs basis parameters along the flat direction as a function of the Higgs trilinear coupling $\kappa_\lambda$.
The gray bands correspond to the $1\sigma$ error bands at the high-luminosity LHC
(see eq.~(\ref{eq:1sigmaunc})).}
\label{fig:flatdirection}
\end{figure}

In fig.~\ref{fig:flatdirection} we plot the values of the single-Higgs coupling parameters as a function of $\kappa_\lambda$ along
the flat direction. It is interesting to notice that a strong correlation is found between the Higgs trilinear coupling $\kappa_\lambda$,
the Higgs contact interaction with gluons $\hat c_{gg}$ and the top Yukawa $\delta y_t$. When we limit the $\kappa_\lambda$
variation to the region $\kappa_\lambda \in [-1, 10]$,
as indicated by the constraints coming from double Higgs production, $\hat c_{gg}$ and $\delta y_t$
vary by an amount comparable with the $1\sigma$ error at the high-luminosity LHC (obtained in a fit without $\kappa_\lambda$). On the other hand, along the flat direction,
the remaining parameters vary by a much lower amount ($\hat c_{\gamma\gamma}$, $\delta c_z$, $\delta y_b$ and $\delta y_\tau$)
or, in some cases, remain almost unchanged ($c_{zz}$, $c_{z\square}$, $\hat c_{z\gamma}$).\footnote{An interesting feature is the
fact that along the flat direction not only $\delta \mu_i^f = 0$, but also the individual production and decay signal strengths
are approximately equal to the SM ones, namely $|\delta \mu_i| < 0.005$, $|\delta \mu^f| < 0.005$ for $|\kappa_\lambda| < 20$.}

It must be stressed that the exact flat direction could in principle be lifted if we include in the signal strengths computation also
terms quadratic in the EFT parameters. The additional terms, however, become relevant only for very large values of
$\kappa_\lambda$,
so that for all practical purposes we can treat the flat direction as exact. Notice moreover that, when the quadratic
terms become important, one must a priori also worry about possible corrections from higher-dimensional operators,
which could become comparable to the square of dimension-$6$ operators.

As we discussed in the previous section, additional observables can provide independent bounds
on the Higgs couplings. In particular some of the strongest constraints come from the measurements of TGC's
and of the $h \rightarrow Z\gamma$ branching ratio. In the fit of the single-Higgs couplings these constraints were enough
to get rid of the large correlation between $c_{zz}$ and $c_{z\square}$ and to improve the bound on $\hat c_{z\gamma}$.
The impact on the global fit including the Higgs trilinear coupling is instead limited.
The reason is the fact that the combination of parameters tested in TGC's (see appendix~\ref{app:trilinear})
and in $h \rightarrow Z\gamma$ are `aligned' with the flat direction, i.e.~they involve couplings whose values along the flat
direction change very slowly (see fig.~\ref{fig:flatdirection}). Although the flat direction is no more exact, even assuming that
the TGC's and $c_{z\gamma}$ can be tested with arbitrary precision, very large deviations in the Higgs self-coupling would still be allowed.

An additional way to probe the flat direction is to compare single-Higgs production rates at different collider energies.
This possibility stems from the fact that the kinematic distributions in Higgs production channels with associated objects
(VBF, $ZH$, $WH$ and $t\bar{t}H$) change in a non-trivial way as a function of the collider energy~\cite{Degrassi:2016wml,Bizon:2016wgr}. As a consequence the impact of
the modification of the Higgs couplings on the production rates shows some dependence on the energy as well.
As one can see from the numerical results reported in appendix~\ref{app:rates}, the dependence of the VBF, $ZH$ and $WH$ rates
on the $c_{zz}$, $c_{z\square}$, $\hat c_{z\gamma}$ and $\hat c_{\gamma\gamma}$ parameters changes as a function of the collider energy
(eqs.~(\ref{eq:sigma_zh}), (\ref{eq:sigma_wh}) and (\ref{eq:sigma_vbf})).
The corrections due to $\kappa_\lambda$ also show a dependence on the energy. In particular the strongest effects are
present in the $t\bar{t}H$ production rate, as can be seen from eq.~(\ref{eq:sigma_NLO}) and the list of coefficients
in table~\ref{tab:cprod}.

The difference in the new physics effects at the
different LHC energies are quite small, so that they do not really allow for an improvement in the fit, taking also into account the fact
that accurate enough predictions will be obtained only for one center of mass energy. Future colliders (as for instance a $33$~TeV
hadron machine) could lead to more pronounced changes in the parameter dependence.\footnote{We thank D.~Pagani for providing
us with the results for the $\kappa_\lambda$ contribution to the inclusive observables at $33$ and $100$~TeV.}
However the improvement achievable with
a combined fit is only marginal. A more efficient way of exploiting higher-energy machines is to look for double Higgs production
which could probe $\kappa_\lambda$ with enough accuracy to make its contributions to single Higgs
processes negligible (assuming that no significant deviation with respect to the SM is found)~\cite{Contino:2016spe}.

\begin{figure}
\centering
\includegraphics[width=.55\textwidth]{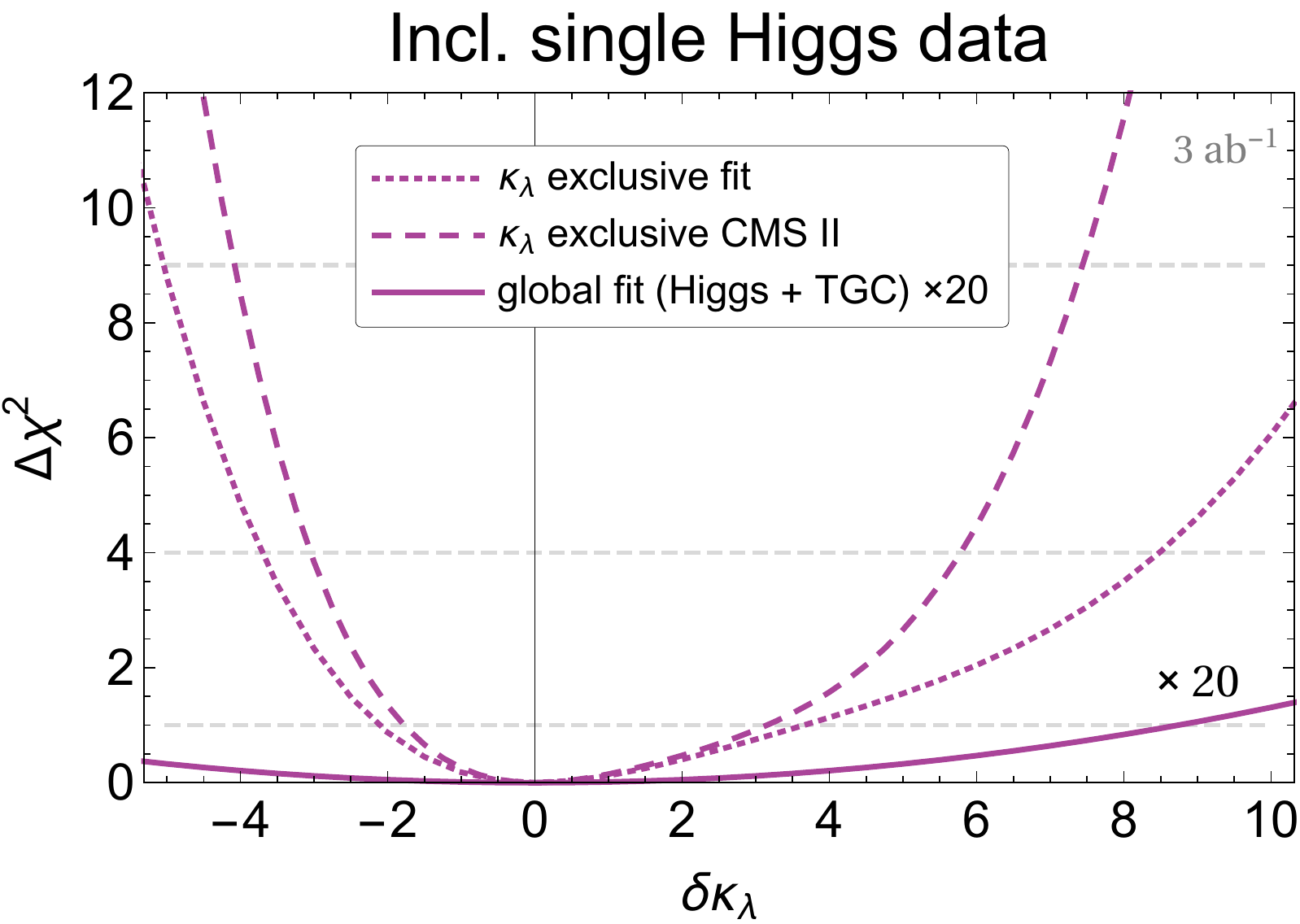}
\caption{ $\chi^2$  as a function of the Higgs trilinear coupling $\kappa_\lambda$ obtained by performing a global fit including
the constraints coming from TGC's measurements and the bound on the $h\to Z\gamma$
decay rate. The results are obtained by assuming an integrated luminosity of $3/{\rm ab}$ at $14$~TeV.
The dotted curve corresponds to the result obtained by setting to zero all the other the Higgs-coupling parameters, while
the solid curve is obtained by profiling and is multiplied by a factor $20$ to improve its visibility.
To compare with previous literature (ref.~\cite{Degrassi:2016wml}), we also display the exclusive fit performed assuming the uncertainty projections
from the more optimistic `Scenario 2' of CMS~\cite{CMS:2013xfa} (dashed curve).}
\label{fig:single_Higgs_kappa}
\end{figure}

To conclude the discussion on the extraction of the Higgs self-coupling, we show in fig.~\ref{fig:single_Higgs_kappa}
the $\chi^2$ obtained from the global fit on single-Higgs observables. The fit also includes the constraints from TGC's and
the bound on the $h\to Z\gamma$ decay rate.\footnote{A full computation of the corrections to the $h\to Z\gamma$
branching ratio due to the Higgs trilinear interaction is not available at present. For this reason we only took into account
the effect of the Higgs wavefunction renormalization, which scales as $\kappa_\lambda^2$ (see appendix~\ref{app:rates}),
and we neglected the additional
corrections linear in $\kappa_\lambda$ which are not known.}
The results have been derived by assuming a $14$~TeV
LHC energy with an integrated luminosity of $3/{\rm ab}$. The dashed curve shows the $\chi^2$ obtained by setting all
the single-Higgs couplings deviations to zero.
One can see that the Higgs self-coupling can be restricted to the interval $\kappa_\lambda \in [-1.1, 4.7]$
at the $1\sigma$ level. To compare with the existing literature, we also show the exclusive fit obtained in the
optimistic `Scenario 2' of CMS (dashed curve), which is in very good agreement with the results of
ref.~\cite{Degrassi:2016wml}.

On the other hand by profiling over the single Higgs couplings we find that the Higgs trilinear coupling
remains basically unconstrained (see solid curve in fig.~\ref{fig:single_Higgs_kappa}).\footnote{Since in our linear
approximation the $\chi^2$ as a function of the single-Higgs couplings
is quadratic the resulting distribution is Gaussian. In this case a profiling procedure gives the same result as a marginalization.}
As expected, even with the inclusion of the TGC's constraints and of the bounds on the $h\to Z\gamma$ decay rate,
an almost flat direction is still present in the fit.

\subsection{Impact of the trilinear coupling on single-Higgs couplings}\label{sec:effects_on_single}

The presence of a flat direction can also have an impact on the fit of the single-Higgs couplings. If we perform a global
fit and we allow $\kappa_\lambda$ to take arbitrary values we also lose predictivity on the single-Higgs EFT parameters.
The effect is more pronounced on the couplings that show larger variations along the flat direction, namely
$\hat c_{gg}$ and $\delta y_t$. A milder impact is found for the $\delta c_z$, $\delta y_b$, $\delta y_\tau$
and $\hat c_{\gamma\gamma}$, whereas $c_{zz}$, $c_{z\square}$ and $\hat c_{z\gamma}$ are almost unaffected, unless
extremely large values of $\kappa_\lambda$ are allowed.

\begin{figure}
\centering
\includegraphics[width=\textwidth]{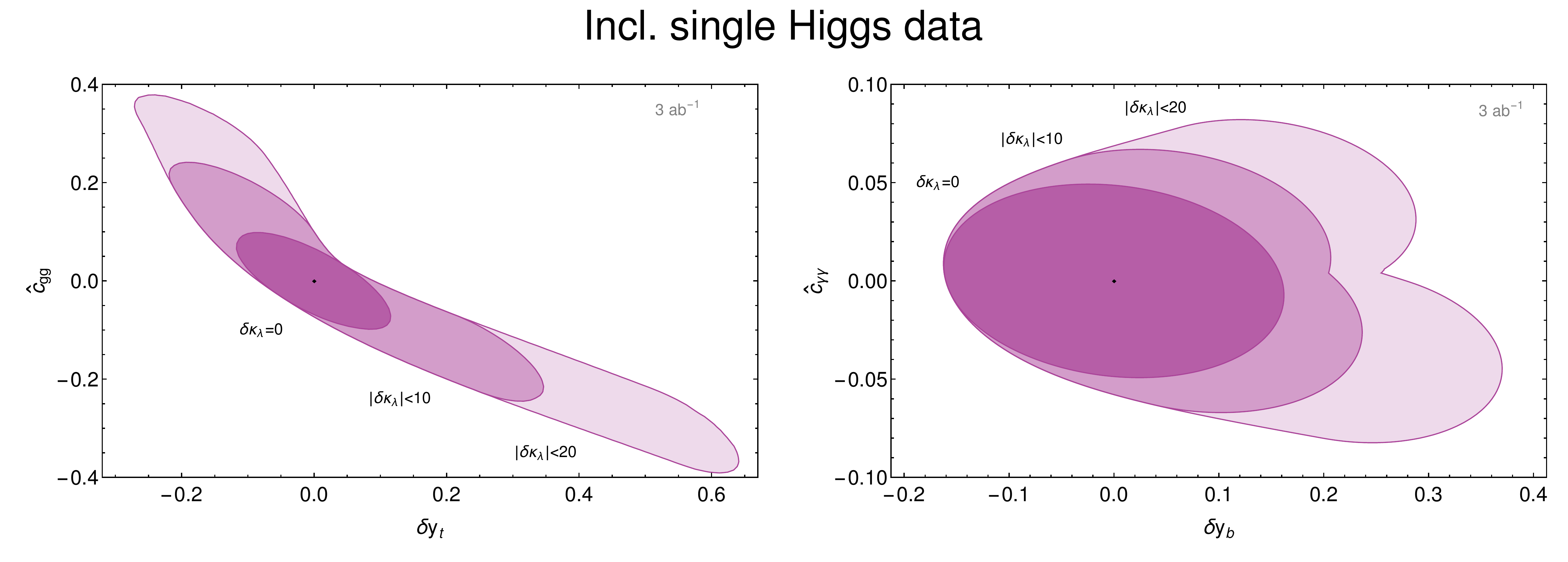}
\caption{Constraints in the planes $(\delta y_t, \hat c_{gg})$ (left panel) and $(\delta y_b, \hat{c}_{\gamma\gamma})$ (right panel)
obtained from a global fit on the single-Higgs processes. The darker regions are obtained by fixing the Higgs trilinear to the
SM value $\kappa_\lambda = 1$, while the lighter ones are obtained through profiling
by restricting  $\delta \kappa_\lambda$ in the ranges $|\delta \kappa_\lambda| \leq 10$ and $|\delta \kappa_\lambda| \leq 20$
respectively. The regions correspond to $68\%$ confidence level (defined in the Gaussian limit corresponding to $\Delta \chi^2 = 2.3$).}
\label{fig:2dcontourplot}
\end{figure}

In fig.~\ref{fig:2dcontourplot} we compare the fit in the $(\delta y_t, \hat c_{gg})$ and $(\delta y_b, \hat c_{\gamma\gamma})$
planes obtained by setting the Higgs trilinear to the SM value ($\delta \kappa_\lambda = 0$), with the results obtained by allowing
$\delta \kappa_\lambda$ to vary in the ranges $|\delta \kappa_\lambda| \leq 10$ and $|\delta \kappa_\lambda| \leq 20$.

In the $(\delta y_t, \hat c_{gg})$ case (left panel of fig.~\ref{fig:2dcontourplot}), there is a strong (anti-)correlation between
the two parameters as we explained in section~\ref{sec:single_Higgs_fit}. When the Higgs self-coupling is included in the fit the
strong correlation is still present. The constraint along the correlated direction becomes significantly weaker,
even if we restrict $\delta \kappa_\lambda$ to the range $|\delta \kappa_\lambda| \leq 10$. The constraint in the orthogonal
direction is instead only marginally affected.

In the case of the $(\delta y_b, \hat c_{\gamma\gamma})$ observables, we find that the $1\sigma$ uncertainty on the
determination of the two parameters is roughly doubled if the Higgs trilinear coupling is allowed to take values up to
$|\delta \kappa_\lambda| \sim 20$.

This above discussion makes clear that a global fit on the single-Higgs observables can not be properly done without
including some assumption on the allowed values of the trilinear self-coupling of the Higgs (see section~\ref{sec:EFT_validity}). If $\kappa_\lambda$ can
sizably deviate from the SM value ($\delta \kappa_\lambda \gtrsim 5$) including it into the fit is mandatory in order to obtain
accurate predictions for the single-Higgs couplings. On the other hand, if we have some theoretical bias that constrains
the Higgs self-coupling modifications to be small ($\delta \kappa_\lambda \lesssim$ few), a restricted fit in which only
the corrections to single-Higgs couplings are included is reliable.

We will see in the following that the situation can drastically change if we include in the fit additional
measurements that can lift the flat direction. In particular we will focus on the measurement
of double Higgs production in the next section and of differential single Higgs distributions in section~\ref{sec:differential}.

\section{Double Higgs production}\label{sec:double_Higgs}

A natural way to extract information about the Higgs self-coupling is to consider Higgs pair production channels. Among this
class of processes, the production mode with the largest cross section~\cite{HHprod},
which we can hope to test with better accuracy at the LHC, is gluon fusion.\footnote{It has been pointed out
in ref.~\cite{Cao:2015oxx} that the $WHH$ and $ZHH$ production modes could provide a good sensitivity to positive
deviations in the Higgs self-coupling (see also refs.~\cite{Barger:1988jk,Moretti:2004wa,Baglio:2012np}). As we will see in the following, the gluon fusion channel is instead more
sensitive to negative deviations. The associated double Higgs production channels could therefore provide useful
complementary information for the determination of $\kappa_\lambda$.
For simplicity we only focus on the gluon fusion channels in the present analysis.
We leave the study of the $VHH$ channels, as well as of the double Higgs production mode in VBF
(see refs.~\cite{Moretti:2004wa,Contino:2010mh,Dolan:2013rja,Dolan:2015zja,Bishara:2016kjn}), for future work.}
Several analyses are available in the literature,
focusing on the various Higgs decay modes. The channel believed to be measurable with the highest
precision is $hh \rightarrow b\overline b \gamma\gamma$~\cite{Baur:2003gp,Grober:2010yv,Contino:2012xk,Baglio:2012np,
Barger:2013jfa,Azatov:2015oxa,He:2015spf,ATL-PHYS-PUB-2014-019,ATL-PHYS-PUB-2017-001}. In spite of the small branching ratio (${\rm BR} \simeq 0.264\%$),
its clean final state allows for high reconstruction efficiency and low levels of backgrounds. In the following we will thus focus on
this channel for our analysis.

Additional final states have also been considered in the literature, in particular
$hh \rightarrow b\overline b b \overline b$~\cite{Baur:2003gpa,Dolan:2012rv,deLima:2014dta,ATL-PHYS-PUB-2016-024},
$hh \rightarrow b\overline bWW^*$~\cite{Dolan:2012rv,Papaefstathiou:2012qe,Baglio:2012np} and
$hh \rightarrow b\overline b\tau^+\tau^-$~\cite{Baur:2003gpa,Dolan:2012rv,Baglio:2012np,Barr:2013tda,Goertz:2014qta}.
All these channels are plagued by much larger backgrounds. In order to extract the signal, one must rely on configurations
with boosted final states and more involved reconstruction techniques, which limit the achievable precision.

The dependence of the double Higgs production cross section on the EFT parameters has been studied in refs.~\cite{Goertz:2014qta,Azatov:2015oxa,Cao:2015oaa,Cao:2016zob}.
It has been shown that a differential analysis taking into account the Higgs pair invariant mass distribution
can help in extracting better bounds on the relevant EFT parameters.

On top of the dependence on $\kappa_\lambda$, double Higgs production is sensitive at leading order to $4$ additional EFT parameters,
namely $\delta y_t$, $\delta y_t^{(2)}$, $\hat c_{gg}$ and $\hat c_{gg}^{(2)}$. The explicit expression of the cross section is given in
appendix~\ref{app:rates}, eq.~(\ref{eq:xhhnewnorm}). As we discussed in section~\ref{sec:eft_parametrization},
in the linear EFT description only $\delta y_t$ and $\hat c_{gg}$ are independent parameters, while the other two correspond to
the combinations given in eq.~(\ref{eq:dependent_params}).
By a suitable cut-and-count analysis strategy, the total SM Higgs pair production cross section is expected to be measured
with a precision $\sim 50\%$ at the high-luminosity LHC~\cite{Azatov:2015oxa}. These estimates are in good agreement with the recent projections by ATLAS~\cite{ATL-PHYS-PUB-2017-001}.

\begin{figure}
\centering
\includegraphics[width=0.475\textwidth]{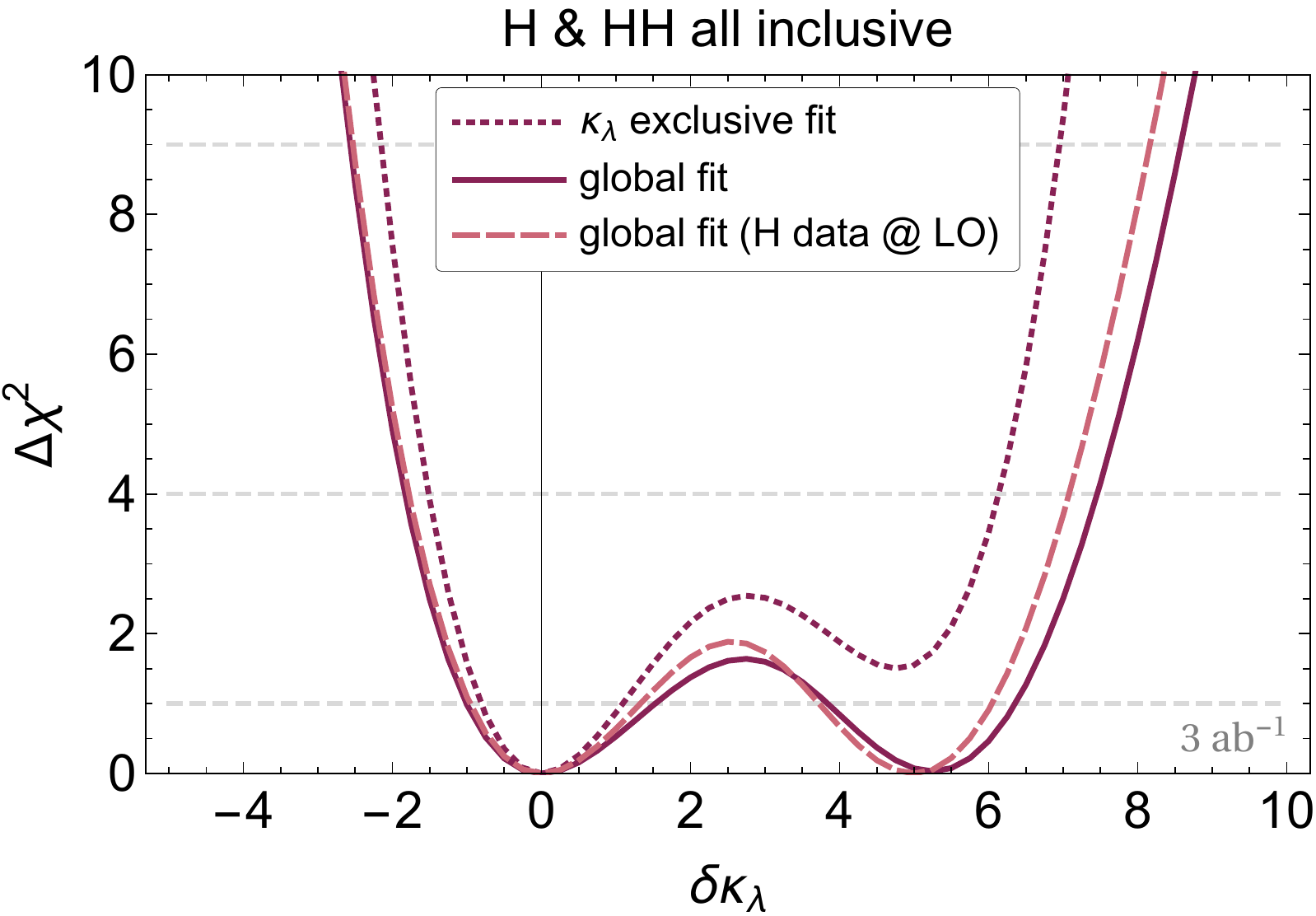}
\hfill
\includegraphics[width=0.475\textwidth]{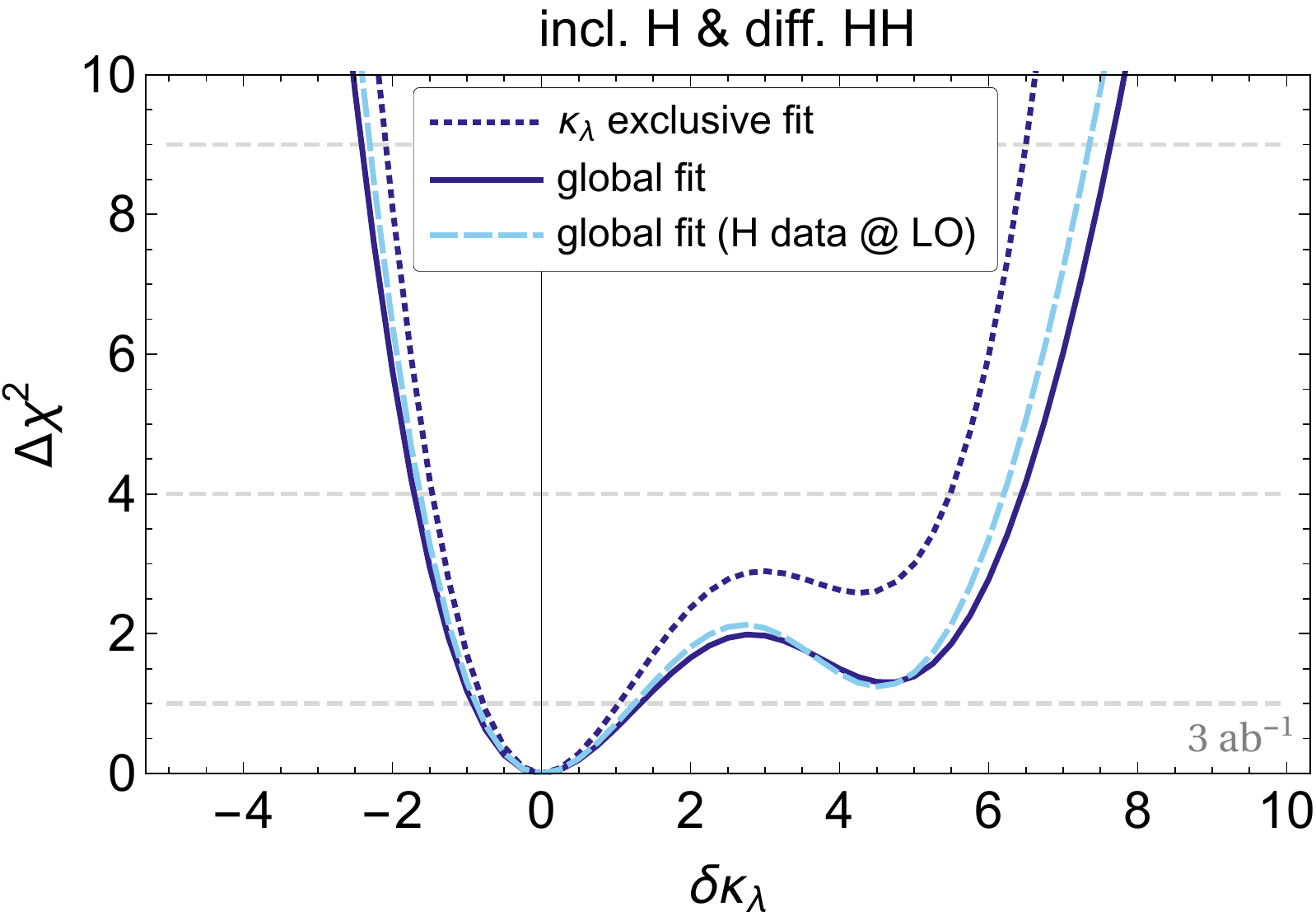}
\caption{\textit{Left:} The solid curve shows the global $\chi^2$ as a function of the corrections to the Higgs trilinear
self-coupling obtained from a fit exploiting inclusive single Higgs and inclusive double Higgs observables.
The dashed line shows the fit obtained by neglecting the dependence on $\delta \kappa_\lambda$ in single-Higgs
observables. The dotted line is obtained by exclusive fit in which all the EFT parameters, except for $\delta \kappa_\lambda$,
are set to zero.
\textit{Right:} The same but using differential observables for double Higgs.}
\label{fig:chi2c3_hh}
\end{figure}

As a first point, we focus on the determination of the trilinear Higgs self-coupling. In the left panel of fig.~\ref{fig:chi2c3_hh}
we show the $\chi^2$ as a function of $\kappa_\lambda$. The solid curve corresponds to the result of
a global fit including single-Higgs and inclusive double-Higgs observables. All the single-Higgs EFT parameters
have been eliminated by profiling.
The dashed curve shows how the fit is modified if we neglect the dependence on $\kappa_\lambda$ in single-Higgs processes.
Finally, the dotted curve is obtained by performing an exclusive fit, in which all the deviations in
single-Higgs couplings are set to zero.

As expected, the measurement of double-Higgs production removes the flat direction that was present in the fit
coming only from single-Higgs observables. The global fit constrains the Higgs trilinear self-coupling to the
intervals $\kappa_\lambda \in [0.0, 2.5] \cup [4.9, 7.4]$ at $68\%$ confidence level and $\kappa_\lambda \in [-0.8, 8.5]$ at $95\%$.
As we can see by comparing the solid and dashed lines in fig.~\ref{fig:chi2c3_hh},
the fit of $\kappa_\lambda$ is almost completely determined by Higgs pair production. This result is expected and is
coherent with the fact that a flat direction involving $\kappa_\lambda$ is present in the single-Higgs fit.
On the other hand if we perform an exclusive fit in which we set to zero all the deviations in single-Higgs couplings,
the determination of the Higgs trilinear self-coupling is significantly modified.
In particular the exclusive fit disfavors large deviations in $\kappa_\lambda$, so that values
$\delta \kappa_\lambda \sim 5$, which were allowed by the global fit, are now excluded at the $1\sigma$ level.
The $95\%$ fit region is also slightly reduced becoming $\kappa_\lambda \in [-0.5, 7.1]$.

It is also interesting to discuss what happens if we include in the fit a differential analysis of double Higgs production.
As shown in ref.~\cite{Azatov:2015oxa}, each new physics effect deforms the Higgs-pair invariant mass distribution in a
different way. Deviations in the Higgs self-coupling mostly affect the threshold distribution, while they have a limited impact
in the high invariant-mass tail. On the contrary $\delta y_{t}$ and $\hat c_{gg}$ modify more strongly the peak and tail
of the distribution. A differential analysis can exploit this different behavior to extract better constraints on the various EFT
coefficients. The fits including the differential information on Higgs pair production are shown in the right panel of fig.~\ref{fig:chi2c3_hh}. 
Sizable positive corrections to $\kappa_\lambda$ are now disfavored even in a global fit. The $1\sigma$ interval is now reduced
to $\kappa_\lambda \in [0.1, 2.3]$, while the $2\sigma$ interval is $\kappa_\lambda \in [-0.7, 7.5]$.

Another aspect worth discussing is the impact of double-Higgs production measurement on the determination
of the single-Higgs couplings. We find that the global fit determines the latter couplings with a precision comparable with 
the one obtained by neglecting the deviations in $\kappa_\lambda$ (see section~\ref{sec:single_Higgs_fit}, eq.~(\ref{eq:1sigmaunc})).
This result may look surprising at a first sight. Double-Higgs measurements at the LHC can only probe the order of magnitude
of the Higgs trilinear self-coupling, so that large deviations from the SM value, $\kappa_\lambda \sim 6$,
will be allowed at the $68\%$ confidence level.
Such big deviations could in turn compensate non-negligible corrections to the single-Higgs measurements (by moving along the
flat direction of the single-Higgs observables fit). The reason why this does not happen is related to the fact that double-Higgs
production is sensitive not only to $\kappa_\lambda$, but also to $\delta y_t$ and $\hat c_{gg}$. Actually, the sensitivity on the
latter two parameters is relatively strong, so that the bounds on $\delta y_t$ and $\hat c_{gg}$ coming from double-Higgs alone
are not much weaker than the ones coming from single-Higgs processes~\cite{Azatov:2015oxa}.
These results hold with the assumption that EW symmetry is linearly realized. We will see in section~\ref{sec:change_uncertainties} how they are modified in the context of a non-linear EFT.

\section{Differential observables}\label{sec:differential}

Up to now we focused on inclusive single Higgs observables, which allowed us to get robust predictions
backed up by the estimates made by the ATLAS and CMS experimental collaborations.
It is however clear that inclusive observables do not maximize the information attainable from the data.
Important additional information can be extracted by exploiting differential single-Higgs distributions.
This can be crucial in our analysis since flat directions are present in the inclusive fit.
Inclusive double-Higgs data is enough to lift this flat direction. Still it leaves a second minimum degenerate
with the SM. Differential information can help removing this degeneracy in addition to improving the precise determination of the Higgs trilinear coupling around the SM.

The exploitation of differential
distributions can help to break the degeneracy  thanks to the fact that the various effective operators
affect the kinematic distributions in different ways.
Consider for instance associated production of a Higgs with a vector boson.
EFT operators that modify the single-Higgs couplings give effects that grow with the centre of mass energy,
hence they mostly affect the high-energy tail of the invariant mass or transverse momentum distributions.
On the contrary, the effect of a modified Higgs trilinear self-coupling is larger near threshold.
This different behavior is the key feature than can allow us to efficiently disentangle the two effects~\cite{Degrassi:2016wml,Bizon:2016wgr}.

The change in the differential single Higgs distributions, in particular in the $WH$, $ZH$, $t\overline t H$ and VBF channels,
as a function of the distortion of the Higgs self-coupling has been
studied in refs.~\cite{Degrassi:2016wml,Bizon:2016wgr}.\footnote{Recently, ref.~\cite{Boselli:2017pef} also computed the impact of the Higgs coupling deviations in the Higgs basis on angular distributions in the four-lepton decay channels of the Higgs boson. We have not included these effects in our analysis.}
In this section we will use these results as a building block to perform a first assessment of the impact of the
differential single-Higgs measurements on the extraction of the Higgs self-interactions and on the global fit of the Higgs couplings.

\subsection{Impact of single-Higgs differential measurements}

In the following we focus our attention on the differential distributions in associated Higgs production channels, $ZH$, $WH$
and $t\overline t H$. We instead neglect the VBF channel, which was found to have a negligible impact on the determination
of the trilinear Higgs coupling in ref.~\cite{Degrassi:2016wml,Bizon:2016wgr}.

For our analysis we consider the differential distributions in the total invariant mass of the processes.
As we discussed in section~\ref{sec:eft_parametrization}, considering high energetic bins in
differential distributions might lead to issues with the validity of the EFT interpretation.
For this reason we only include in our analysis bins with an invariant mass up to three times the threshold energy for the
various channels,
which corresponds to $\sim 600$~GeV for associated production with a gauge boson and to
$\sim 1.4$~TeV for $t\bar{t}H$. The numerical LO predictions of the $ZH$ and $WH$ cross sections
in each bin as a function of the single-Higgs EFT parameters are given in appendix~\ref{app:rates}, while the signal strength for $t\bar{t}H$ is instead modified at LO in an energy-independent way. Concerning the loop-induced effect of $\kappa_\lambda$ on the invariant mass distributions of the $ZH$, $WH$, and $t\bar{t}H$ cross-sections,
only the $13$~TeV results are known~\cite{Degrassi:2016wml}. Therefore we use this center of mass energy for
our numerical study. We however expect that our results provide a fair assessment of the precision achievable at the
$14$~TeV high-luminosity LHC, since the differences with respect to the $13$~TeV case should not be very large.

For our numerical analysis we estimate the statistical and systematic uncertainties from the high-luminosity-LHC ATLAS projections~\cite{ATLAS:projections}. A comprehensive analysis of the uncertainties at the differential level is beyond the scope of our study. We therefore adopt some simplified assumptions to provide a first assessment of the benefit of including differential distributions in our global fit of single-Higgs observables.
In order to evaluate the dependence of our results on the experimental accuracy
we consider two different procedures to estimate the uncertainties.
In the first, more optimistic procedure, the systematic uncertainty is assumed to be the same in all the invariant
mass bins, whereas the statistical uncertainty is rescaled according to the expected number of events in each bin.
In the second, more pessimistic estimate, we extract the uncertainty for each bin by rescaling the total
experimental error according to the expected number of events in each bin. In this way we are effectively
inflating the systematic errors assuming that they degrade as the statistical ones in bins with fewer events.
The uncertainties for the two scenarios are reported in tables~\ref{tab:differential2_error} and~\ref{tab:differential_error}.

Notice that the invariant mass of some processes is not directly accessible experimentally, since the event kinematics
can not be fully reconstructed. We nevertheless use it for our analysis for simplicity.
As a cross check, we verified that performing the analysis with transverse momentum binning does not significantly
modify the results of the fit.
Since our estimates of the experimental uncertainties and our analysis strategy are quite crude, we do not expect
our numerical results to be fully accurate. They must instead be interpreted as rough estimates which can however give
an idea of the discriminating power that we could expect by the exploitation of differential single-Higgs distributions.

As a first step we consider the impact on the determination of single-Higgs couplings. Including the differential information
in the fit helps in reducing the correlation between $c_{zz}$ and $c_{z\square}$. The overall change in the fit is however
small and the $1\sigma$ intervals are nearly unchanged with respect to the ones we obtained in the inclusive analysis
(see eq.~(\ref{eq:1sigmaunc_tgcconstr})).

\begin{figure}
\includegraphics[width=0.475\textwidth]{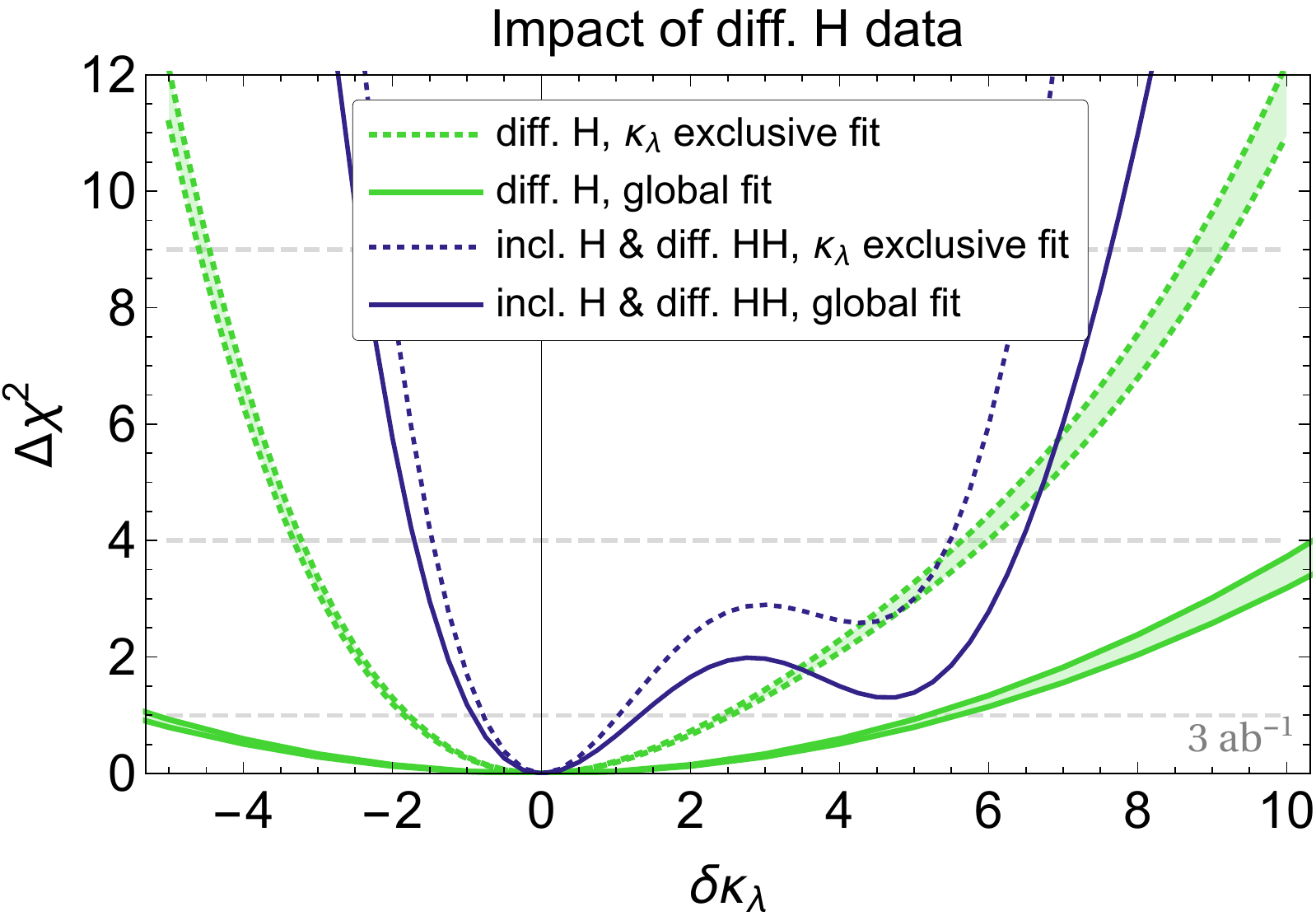}
\hfill
\includegraphics[width=0.475\textwidth]{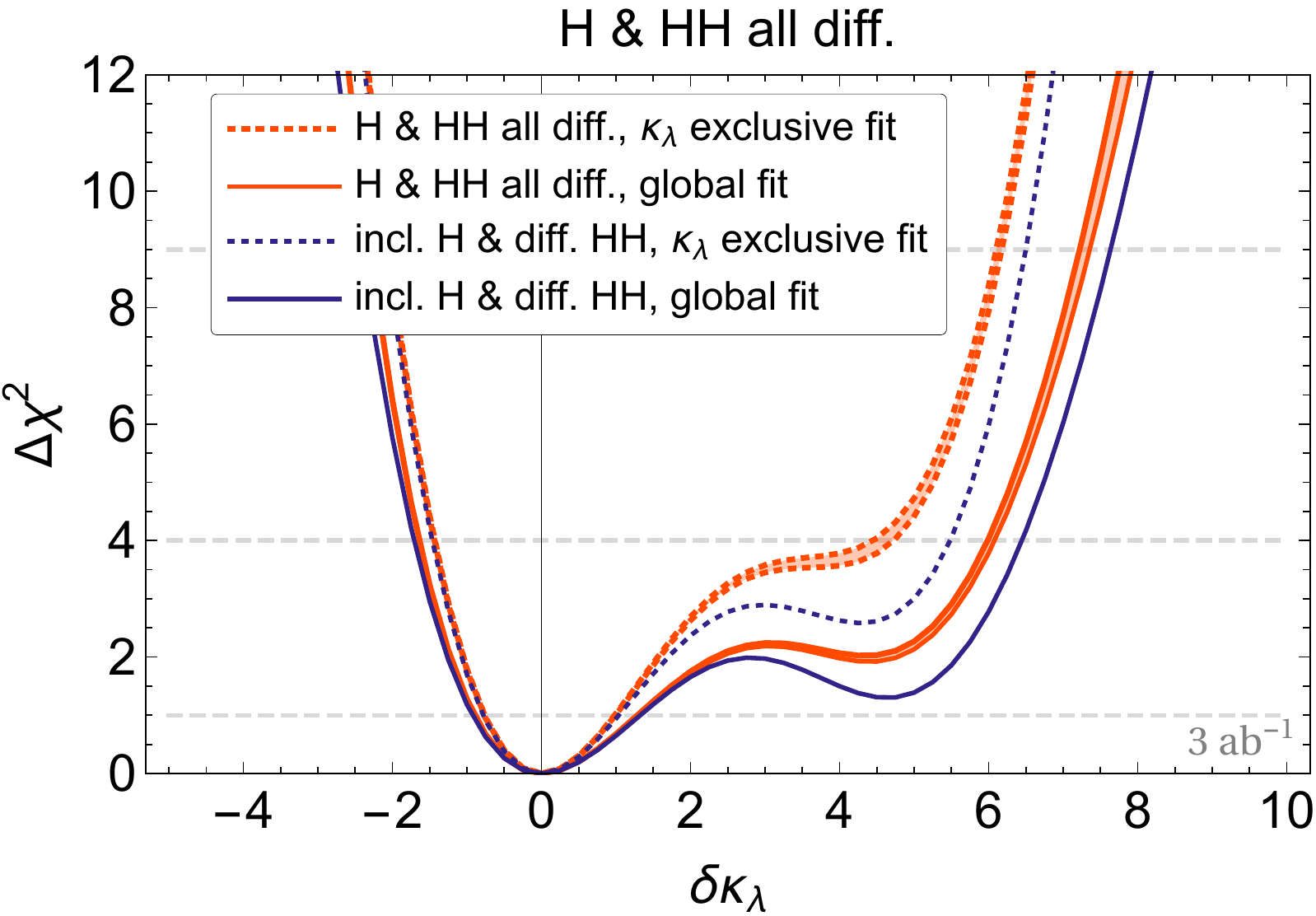}
\caption{\textit{Left:} $\chi^2$ as a function of the Higgs trilinear self-coupling. The green bands are obtained from the differential analysis on single-Higgs
observables and are delimited by the fits corresponding to the optimistic and pessimistic estimates of the experimental
uncertainties.
The dotted green curves correspond to a fit performed exclusively on $\delta \kappa_\lambda$ setting to zero all the other parameters,
while the solid green lines are obtained by a global fit profiling over the single-Higgs coupling parameters.
\textit{Right:} The red lines show the fits obtained by a combination of single-Higgs and double-Higgs differential observables.
In both panels the dark blue curves are obtained by considering only double-Higgs differential observables
and coincide with the results shown in fig.~\ref{fig:chi2c3_hh}.
}
\label{fig:chi2c3_differential}
\end{figure}

More interesting results are instead obtained when we focus on the extraction of the Higgs trilinear self-coupling.
We find that differential distributions are able to lift the flat direction we found in the inclusive single-Higgs observables fit.
The solid green lines in fig.~\ref{fig:chi2c3_differential} show the $\chi^2$ obtained in a global fit
on single-Higgs observables including the differential information from associated production modes. The two lines correspond
to the `optimistic' and 'pessimistic' assumptions on the experimental uncertainties.
Through this procedure one could constrain the Higgs trilinear coupling to the interval $|\delta \kappa_\lambda| \lesssim 5$ at the $1\sigma$ level.
An exclusive fit, in which all the single-Higgs couplings deviations are set to zero, gives a range
$\kappa_\lambda \in [-0.8, 3.5]$ at $1\sigma$ and $\kappa_\lambda \in [-2, 7]$ at $2\sigma$ (dotted green lines),
which is significantly smaller than the one obtained through a global fit, as can be seen by comparing with the solid lines in fig.~\ref{fig:chi2c3_differential}.

The results in fig.~\ref{fig:chi2c3_differential} show that
in a global fit the impact of differential single-Higgs measurements on the extraction of $\kappa_\lambda$ is weaker
than the one of differential double-Higgs production. This can be clearly seen by comparing the solid green lines
with the solid dark blue curve which represent the $\chi^2$ coming from double Higgs
measurements (this curve coincides with the results shown on the right panel of fig.~\ref{fig:chi2c3_hh}).
Nevertheless, combining the single-Higgs differential information with the double-Higgs fit helps in testing large positive deviations
in $\kappa_\lambda$, increasing the $\chi^2$ value for values $\delta \kappa_\lambda \sim 5$. This improvement can
be seen on the right panel of fig.~\ref{fig:chi2c3_differential} (solid curves).

Differential single-Higgs measurements have a significantly more relevant role in exclusive fits in which the single-Higgs
parameters are set to zero. One can see in the left panel of fig.~\ref{fig:chi2c3_differential} that the sensitivity of
the single-Higgs differential fit (dotted blue line) is comparable with the one of double-Higgs measurements, especially
for positive deviations in $\kappa_\lambda$. Combining single-Higgs and double-Higgs information provides a
good improvement in the fit, in particular at the $2\sigma$ level, as can be seen in the right panel of
fig.~\ref{fig:chi2c3_differential} (dotted lines).

\subsection{Robustness of the fits}\label{sec:change_uncertainties}

As a final point we want to discuss how much the determination of the Higgs trilinear self-interaction and of the
single-Higgs couplings depends on the experimental accuracy and on the theoretical assumptions underlying
the EFT parametrization.

\begin{figure}
\centering
\includegraphics[width=0.475\textwidth]{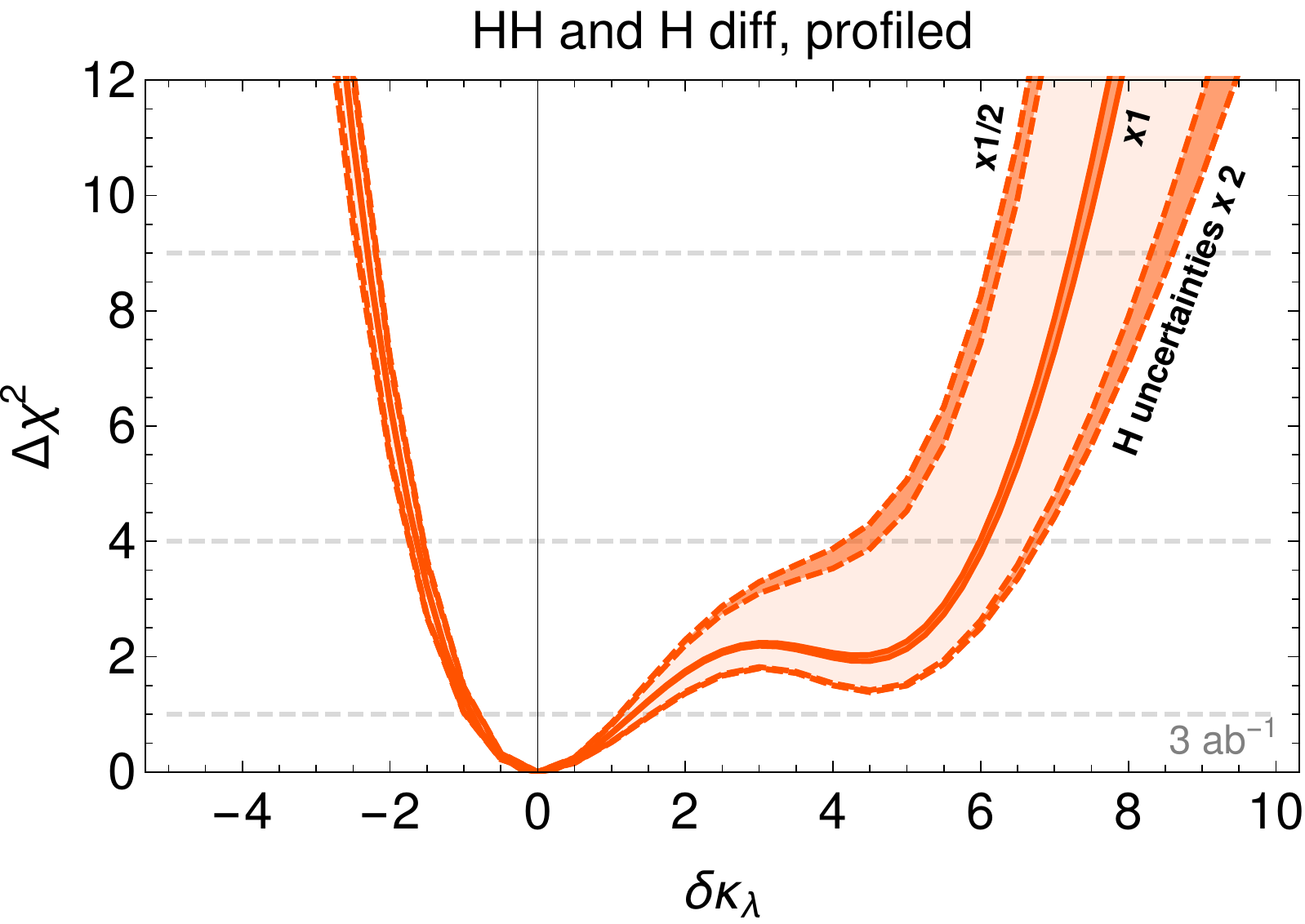}
\hfill
\includegraphics[width=0.475\textwidth]{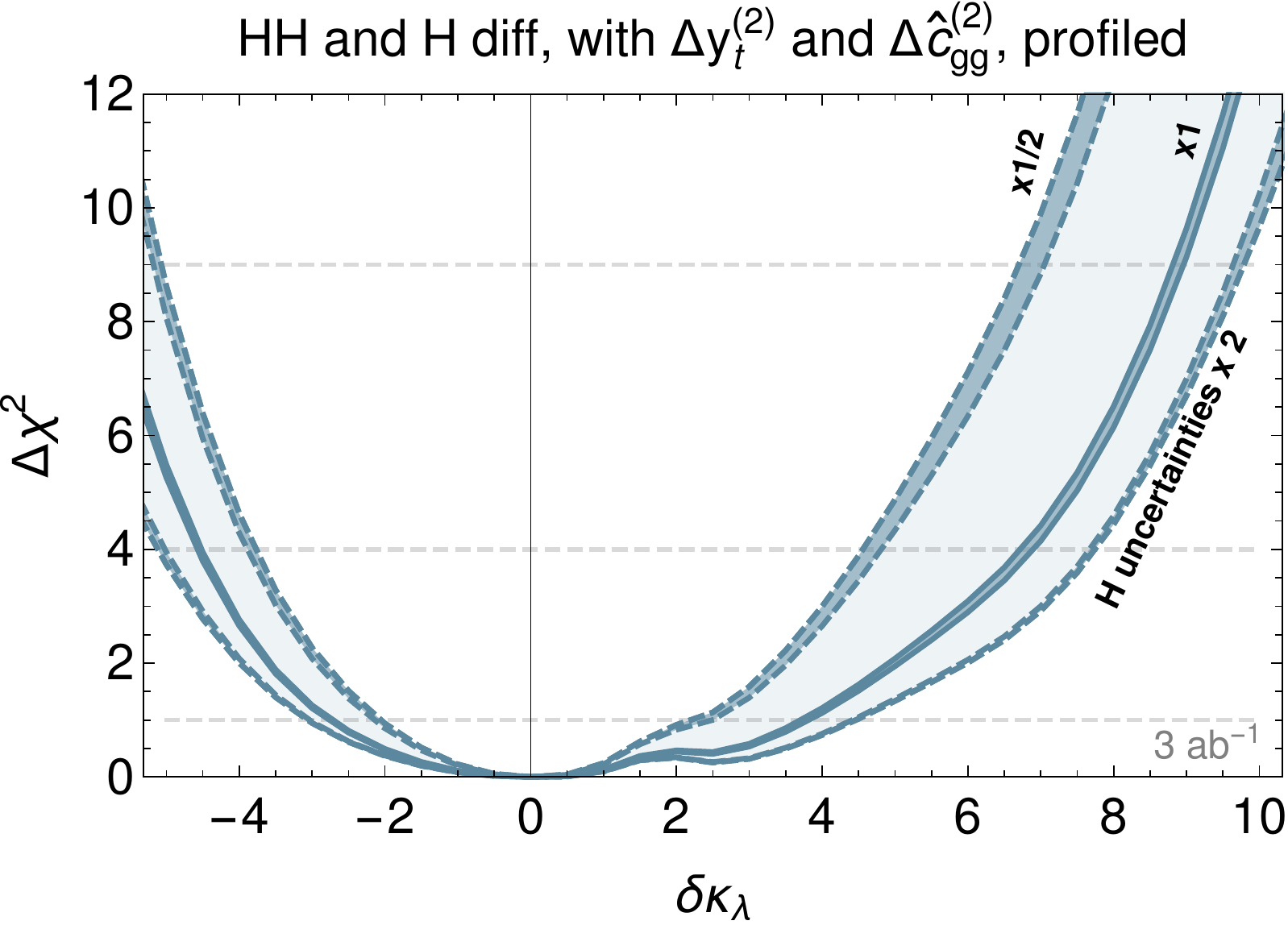}
\caption{Band of variation of the global fit on the Higgs self-coupling obtained by rescaling the single-Higgs measurement
uncertainties by a factor in the range $x \in [1/2, 2]$. The lighter shaded bands show the full variation of the fit due to the rescaling.
The darker bands show how the fits corresponding to the `optimistic' and `pessimistic'
assumptions on the systematic uncertainties (compare fig.~\ref{fig:chi2c3_differential}) change for $x = 1/2, 1, 2$.
The left panel shows the fit in the linear Lagrangian, while the right panel corresponds to the non-linear case
in which $\Delta y_f^{(2)}$ and $\Delta \hat{c}_{gg}^{(2)}$ are treated as independent parameters.
}
\label{fig:rescaling}
\end{figure}

In the left panel of fig.~\ref{fig:rescaling} we show how the fit on $\kappa_\lambda$ changes if we rescale the errors on
single-Higgs measurements by a factor in the range $[1/2, 2]$. One can see that the $\chi^2$ function around the
SM point $\delta \kappa_\lambda = 0$ is not strongly affected, so that the $1\sigma$ region is only
mildly modified. Large positive deviations from the SM can instead be probed with
significantly different accuracy. In particular the $2\sigma$ region is enlarged to $\kappa_\lambda \in [-0.8, 7.7]$
if we double the uncertainties, whereas it shrinks to $\kappa_\lambda \in [-0.5, 5.3]$ if we reduce the errors by a factor $1/2$.

A second point worth investigating is how the fit changes if we modify the assumptions on the EFT parametrization.
As an illustrative example we analyze a scenario in which the EFT Lagrangian has a non-linear form, i.e.~the expansion
in Higgs powers breaks down. As we already discussed in section~\ref{sec:EFT_validity},
in this case operators containing Higgs fields can not be
tested any more in precision measurements not involving the Higgs. A fully consistent fit should thus include all possible operators
and not just the restricted basis we defined in eqs.~(\ref{eq:coupl_def}) and ~(\ref{eq:dependent_params}).
Performing such analysis is beyond the
scope of the present paper. For illustration we restrict our attention only to two effective operators, $h^2 G_{\mu\nu} G^{\mu\nu}$
and $h^2 \overline t t$, whose impact on Higgs pair production via gluon fusion was studied in ref.~\cite{Azatov:2015oxa}.

In the linear EFT Lagrangian the $h^2 G_{\mu\nu} G^{\mu\nu}$
and $h^2 \overline t t$ operators are connected to single-Higgs couplings (see eq.~\ref{eq:dependent_params}).
Treating them as independent operators amounts to including the $\delta y_f^{(2)}$ and $\delta\hat{c}_{gg}^{(2)}$ parameters
as free quantities in our fits. For convenience we introduce two new parameters that encode the deviations
of $\delta y_f^{(2)}$ and $\hat{c}_{gg}^{(2)}$ from the linear Lagrangian relations:
\begin{equation}
\Delta y_f^{(2)} \equiv \delta y_f^{(2)} - (3 \delta y_u - \delta c_z)
\,, \qquad \quad
\Delta\hat{c}_{gg}^{(2)} \equiv \hat{c}_{g g}^{(2)} - \hat{c}_{gg}\,.
\end{equation}

To understand the impact of $\Delta y_f^{(2)}$ and $\Delta \hat{c}_{gg}^{(2)}$ on the global fit, we give in the following equation
the $1\sigma$ intervals for the Higgs couplings in the linear and non-linear scenarios
\begin{equation}
\begin{array}{c@{\hspace{7em}}c}
\rule[-.75em]{0pt}{.5em}{\rm \textit{Fit\ with}}\ \Delta y_f^{(2)} = \Delta\hat{c}_{gg} = 0 & {\rm \textit{Global\ fit}}\\
\left(
\begin{array}{c}
\hat c_{gg} \\ \delta c_z \\ c_{zz} \\ c_{z\square} \\ \hat c_{z\gamma} \\ \hat c_{\gamma\gamma}\\ \delta y_t \\ \delta y_b \\ \delta y_\tau \\
\delta \kappa_\lambda
\end{array}
\right)
= \pm
\left(
\begin{array}{cc}
0.06\\
0.04\\
0.04\\
0.02\\
0.09\\
0.03\\ 
0.06\\
0.07\\
0.11\\
1.0
\end{array}
\right)\,,
&
\left(
\begin{array}{c}
\hat c_{gg} \\ \delta c_z \\ c_{zz} \\ c_{z\square} \\ \hat c_{z\gamma} \\ \hat c_{\gamma\gamma}\\ \delta y_t \\ \delta y_b \\ \delta y_\tau \\
\delta \kappa_\lambda\\ \Delta y_f^{(2)} \\ \Delta\hat{c}_{gg}^{(2)} 
\end{array}
\right)
= \pm
\left(
\begin{array}{cc}
0.07\\
0.04\\
0.04\\
0.02\\
0.09\\
0.03\\ 
0.08\\
0.08\\
0.11\\
4.1\\
\rule{0pt}{1.1em} 0.29\\
\rule{0pt}{1.05em} 0.45
\end{array}
\right)\,.
\end{array}
\end{equation}
One can see that the non-linear fit mostly affects the determination of $\kappa_\lambda$, whose precision significantly degrades.
The impact on the determination of single-Higgs couplings is instead quite limited and is due to the fact that a weaker
constraint on the Higgs self-interaction allows to move along the flat direction in the single-Higgs global fit. Indeed we find that
the $1\sigma$ intervals for $\hat c_{gg}$, $\delta y_t$ and $\delta y_b$ are slightly larger in the non-linear scenario. The differences
are however only marginally relevant.

To better quantify how the determination of $\kappa_\lambda$ changes in the non-linear case, we show the $\chi^2$
obtained in the global fit in the right panel of fig.~\ref{fig:rescaling}. The $1\sigma$ band in this case becomes
$\kappa_\lambda \in [-2, 5]$. We also show how the fit depends on the precision in the measurement of the single-Higgs
observables. One can see that a reduction by a factor $1/2$ of the uncertainties on single-Higgs measurements could help
significantly in improving the determination of $\kappa_\lambda$, reducing the $1\sigma$ band by $\sim 40\%$.

\section{Conclusions}\label{sec:conclusions}

The possibility of exploiting single-Higgs production channels at hadron colliders to extract information about the Higgs trilinear
self-coupling has been recently put forward in the
literature~\cite{Gorbahn:2016uoy,Degrassi:2016wml,Bizon:2016wgr}. The available results are quite encouraging.
They show that the new analysis strategy could be competitive with the study of double-Higgs production,
which is usually considered the best way to probe the Higgs self-interactions.

The analyses performed so far, however, limited their focus to scenarios in which the only deformation of the SM Lagrangian
is a modification of the Higgs potential. This assumption significantly restricts the
realm of theories for which the new results are valid. Indeed, in a vast class of new-physics models, corrections to the
Higgs trilinear coupling are not generated alone and additional deviations in the other Higgs interactions are
simultaneously present. Since the Higgs self-coupling only affects at next-to-leading order the single-Higgs rates,
its effects can be easily overwhelmed by even small modifications of the single-Higgs couplings. In this more generic
situation a global analysis, taking into account deviations in all the Higgs couplings simultaneously, is essential to
fully assess the achievable accuracy. The main aim of the present paper has been to perform such analysis.
The computations of refs.~\cite{Gorbahn:2016uoy,Degrassi:2016wml,Bizon:2016wgr} are an essential building block that can be directly implemented in a global fit with all the parameters affecting the Higgs couplings turned on simultaneously.

For definiteness we studied deformations of the SM Lagrangian given by dimension-$6$ effective operators in the
SMEFT framework. In particular, in addition to deviations in the Higgs self-coupling, we considered distortions of the
single-Higgs couplings due to a set of $9$ operators that can not be tested with $\%$ precision in measurements not involving the Higgs.
In the Higgs basis these deformations are encoded in the coefficients $\delta c_z$, $c_{zz}$, $c_{z\square}$,
$\hat c_{z\gamma}$, $\hat c_{\gamma\gamma}$ and $\hat c_{gg}$ which correspond to deformations of the Higgs couplings
to gauge bosons, and $\delta y_t$, $\delta y_b$ and $\delta y_\tau$ controlling deformations of the Yukawa's.

To derive our numerical results we considered the high-luminosity LHC upgrade ($14$~TeV center of mass energy and
$3/$ab integrated luminosity) and we estimated the precision on single-Higgs measurements through a benchmark
derived from the ATLAS and CMS projections~\cite{ATLAS:projections,CMS:2013xfa} (see table~\ref{tab:errors}).
Moreover we assumed that the central values of the future experimental measurements will coincide with the SM predictions.

We found that, if only inclusive single-Higgs observables are considered, a global fit involving the $10$ free parameters has an (almost) exact
flat direction. The flat direction is mostly aligned along the Higgs self-coupling $\kappa_\lambda$, the top Yukawa
$\delta y_t$ and the contact interaction with gluons $\hat c_{gg}$, with minor components along $\delta c_z$, $\delta y_b$
and $\delta y_\tau$ (see fig.~\ref{fig:flatdirection}). The inclusion of trilinear gauge couplings measurements can only partially lift
the flat direction. Very large deviations in $\kappa_\lambda$ are however still allowed,
so that the Higgs self-interaction remains practically unbounded (see fig.~\ref{fig:single_Higgs_kappa}). This result clearly
shows that the bounds obtained by an exclusive fit including only $\kappa_\lambda$ ($\kappa_\lambda \in [-1.1, 4.7]$
at the $1\sigma$ level) must be interpreted with great care and are fully valid only in very specific BSM scenarios.

Large deviations in the Higgs self-coupling can also have a back-reaction on the extraction of single-Higgs couplings.
As can be seen from fig.~\ref{fig:2dcontourplot}, if large corrections, $|\delta \kappa_\lambda| \sim 10$, are allowed, the precision
in the determination of the single-Higgs couplings is significantly degraded. This results shows the necessity
of including in the global fit additional observables which could resolve the flat direction.

We explored two possible extensions of the fitting procedure, namely the inclusion of double-Higgs production via gluon fusion and
the use of differential measurements in the associated single-Higgs production channels $WH$, $ZH$ and $t\overline t H$.

As expected, an inclusive double Higgs production measurement can efficiently remove the flat direction,
constraining the Higgs trilinear coupling to the range $\kappa_\lambda \in [0.0, 2.5] \cup [4.9, 7.4]$ at the $1\sigma$
confidence level (see fig.~\ref{fig:chi2c3_hh}). Furthermore, differential double-Higgs distributions can
provide additional help to probe large
positive deviations in the Higgs trilinear. In particular they can be used to test the additional best fit point at $\kappa_\lambda \sim 6$
and to reduce the $2\sigma$ fit range (see right panel of fig.~\ref{fig:chi2c3_hh}).
When differential double-Higgs measurements are included, the constraint on the Higgs trilinear coupling becomes
$\kappa_\lambda \in [0.1, 2.3]$ at the $1\sigma$ confidence level, which is strong enough
to ensure that the back-reaction on the single-Higgs couplings fit is almost negligible at the high-luminosity LHC.
This result proves that neglecting the contributions from $\kappa_\lambda$ when performing a fit on single Higgs couplings
is a sensible procedure, even in BSM scenarios that can lead to ${\cal O}(1)$ deviations in the Higgs self-interactions.

The measurement of the differential distributions in the associated Higgs production channels can also help in determining
the Higgs self-coupling. In the present paper we performed a preliminary analysis with a simplified treatment
of the experimental and theory uncertainties. We found that an exclusive fit on $\kappa_\lambda$ can provide order one sensitivity
($\kappa_\lambda \in [-1, 3]$ at $1\sigma$), roughly comparable with the one achievable through double-Higgs measurements
(see fig.~\ref{fig:chi2c3_differential}).
On the other hand, in a global analysis, including deviations in single-Higgs couplings, the sensitivity on $\kappa_\lambda$
is strongly reduced and only large deviations $|\delta \kappa_\lambda| \gtrsim 5$ can be probed.
Nevertheless, also in this case single-Higgs differential observables can be useful. Combining them with double-Higgs
measurements can significantly help to constrain large positive corrections to the Higgs trilinear. To fully evaluate the
impact of the differential observables a more careful analysis strategy, together with a detailed assessment of the experimental
uncertainties, would be needed. We leave this subject for future work.

Another important aspect we investigated is the dependence of our results on the experimental uncertainties and on the assumptions
underlying the EFT parametrization. As shown in the left panel of fig.~\ref{fig:rescaling}, a naive rescaling of all the
experimental uncertainties in single-Higgs production affects only mildly the bounds on negative contributions
to $\kappa_\lambda$, but has a major impact on the constraints on positive corrections
(in particular at the $2\sigma$ confidence level).

The assumptions on the EFT description can also strongly modify the determination of $\kappa_\lambda$. As an illustrative example
we considered a non-linear EFT Lagrangian in which the double-Higgs couplings to gluons and to tops are treated as independent
parameters. This change affects almost exclusively the precision on the Higgs self-coupling, which is reduced by roughly
a factor $3$ (right panel of fig.~\ref{fig:rescaling}). On the contrary, the global fit on the single-Higgs couplings is much more
stable and only the determination of $\hat c_{gg}$ and $\delta y_t$ becomes marginally worse.

\section*{Acknowledgements}
We thank R.~Contino, G.~Degrassi, A.~Falkowski, U.~Haisch, A.~Nisati, D.~Pagani and, especially, A.~Pomarol for several useful discussions and suggestions.
We acknowledge G.~Degrassi, P.~P.~Giardino, F.~Maltoni, and D.~Pagani, and A.~Nisati for comments on the manuscript.
We are also grateful to A.~Falkowski and R.~Rattazzi
for sharing with us the results of their work~\cite{Falkowski} before publication.
C.G. is supported by the European Commission through the Marie Curie Career Integration Grant 631962 and by the Helmholtz Association. M.R. is supported by la Caixa, Severo Ochoa grant program. G.P., M.R. and T.V. are supported by the Spanish Ministry MEC under grants FPA2014-55613-P and FPA2011-25948, by the Generalitat de Catalunya grant 2014-SGR-1450 and by the Severo Ochoa excellence program of MINECO (grant SO-2012-0234). We thank the Collaborative Research Center SFB676 of the Deutsche Forschungsgemeinschaft (DFG), “Particles, Strings and the Early Universe”, for support.

\appendix

\section{Higgs production and decay rates in the EFT framework}\label{app:rates}

In this appendix we report the expressions for the production and decay rates of the Higgs boson as a function
of the EFT parameters. The numerical results have been obtained at LO through \texttt{Feynrules}~\cite{Alloul:2013bka} and \texttt{MadGraph}~\cite{Alwall:2014hca} by using the
model ``Higgs effective Lagrangian''~\cite{Alloul:2013naa}.

We start by listing the dependence on the single-Higgs couplings deformations ($\delta c_z$, $c_{zz}$, $c_{z\square}$,
$\hat c_{z\gamma}$, $\hat c_{\gamma\gamma}$, $\hat c_{gg}$, $\delta y_t$, $\delta y_b$, $\delta y_\tau$).
The modification of the total cross sections for associated production ($ZH$ and $WH$) and VBF depend on the collider energy.
The results at $7$, $8$, $13$, $14$, $33$ and $100$ $\tev{}$ are given by
\begin{equation}
  \frac{\sigma_{ZH}}{\sigma_{ZH}^{\text{SM}}} = 1+ \delta c_z\left(
\begin{array}{c}
 2.0 \\
 2.0 \\
 2.0 \\
 2.0 \\
 2.0 \\
 2.0 \\
\end{array}
\right)+
c_{z \square}\left(
\begin{array}{c}
 7.6 \\
 7.8 \\
 8.3 \\
 8.4 \\
 9.1 \\
 10.0 \\
\end{array}
\right)+
c_{zz}\left(
\begin{array}{c}
 3.4 \\
 3.4 \\
 3.5 \\
 3.6 \\
 3.7 \\
 4.0 \\
\end{array}
\right)-
\hat{c}_{z\gamma}\left(
\begin{array}{c}
 0.060 \\
 0.061 \\
 0.067 \\
 0.068 \\
 0.077 \\
 0.086 \\
\end{array}
\right)-
\hat{c}_{\gamma\gamma}\left(
\begin{array}{c}
 0.028 \\
 0.028 \\
 0.030 \\
 0.032 \\
 0.034 \\
 0.037 \\
\end{array}
\right)\,,\label{eq:sigma_zh}
\end{equation}

\begin{equation}
  \frac{\sigma_{WH}}{\sigma_{WH}^{\text{SM}}} = 1+ \delta c_z\left(
\begin{array}{c}
 2.0 \\
 2.0 \\
 2.0 \\
 2.0 \\
 2.0 \\
 2.0 \\
\end{array}
\right)+
c_{z {\scriptscriptstyle \square}}\left(
\begin{array}{c}
 9.3 \\
 9.4 \\
 10.0 \\
 10.1 \\
 11.1 \\
 12.1 \\
\end{array}
\right)+
c_{zz}\left(
\begin{array}{c}
 4.4 \\
 4.4 \\
 4.6 \\
 4.6 \\
 5.0 \\
 5.3 \\
\end{array}
\right)-
\hat{c}_{z\gamma}\left(
\begin{array}{c}
 0.082 \\
 0.084 \\
 0.094 \\
 0.095 \\
 0.110 \\
 0.126 \\
\end{array}
\right)-
\hat{c}_{\gamma\gamma}\left(
\begin{array}{c}
 0.044 \\
 0.045 \\
 0.048 \\
 0.049 \\
 0.054 \\
 0.060 \\
\end{array}
\right)\,,\label{eq:sigma_wh}
\end{equation}

\begin{equation}
  \frac{\sigma_{VBF}}{\sigma_{VBF}^{\text{SM}}} = 1+ \delta c_z\left(
\begin{array}{c}
 2.0 \\
 2.0 \\
 2.0 \\
 2.0 \\
 2.0 \\
 2.0 \\
\end{array}
\right)-
c_{z \square}\left(
\begin{array}{c}
 2.2 \\
 2.2 \\
 2.5 \\
 2.5 \\
 3.0 \\
 3.7 \\
\end{array}
\right)-
c_{zz}\left(
\begin{array}{c}
 0.81 \\
 0.83 \\
 0.89 \\
 0.90 \\
 1.04 \\
 1.27 \\
\end{array}
\right)+
\hat{c}_{z\gamma}\left(
\begin{array}{c}
 0.029 \\
 0.030 \\
 0.033 \\
 0.034 \\
 0.041 \\
 0.051 \\
\end{array}
\right)+
\hat{c}_{\gamma\gamma}\left(
\begin{array}{c}
 0.0113 \\
 0.0117 \\
 0.0129 \\
 0.0131 \\
 0.0156 \\
 0.0193 \\
\end{array}
\right)\,,\label{eq:sigma_vbf}
\end{equation}
where we employ the VBF cross section definition of ref.~\cite{Falkowski:2015fla},
namely we apply the following cuts on the two forward jets: $p_{T,j}>20\,\gev{}$, $|\eta_j|<5$, and $m_{jj}>250\,\gev{}$.

The cross sections of the gluon fusion and $t\overline{t}H$ production modes are instead modified in an energy-independent
way~\cite{Falkowski:2015fla}.
This is a consequence of the fact that at LO the gluon fusion energy scale is fixed by the Higgs bosons on-shell condition
and is therefore $\sqrt{s}$ independent, while the modification of $t\overline{t}H$ is simply due to a rescaling of the top Yukawa.
\begin{alignat}{2}
&\frac{\sigma_{\text{ggF}}}{\sigma_{\text{ggF}}^{\text{SM}}} &&= 1+ 2 \hat{c}_{gg} + 2.06 \delta y_t - 0.06 \delta y_b\,,\\
&\frac{\sigma_{\text{ttH}}}{\sigma_{\text{ttH}}^{\text{SM}}} &&= 1+2\delta y_t\,.
\end{alignat}

The modifications of the decay widths are given by~\cite{Falkowski:2015fla}
\begin{alignat}{3}
&\frac{\Gamma_{\gamma\gamma}}{\Gamma_{\gamma\gamma}^{\text{SM}}} &= & 
\;1 + 2.56\,\delta c_z +2.13 \, c_{z\square } + 0.98\, c_{zz} - 0.066 \hat{c}_{z\gamma} - 2.46\, \hat{c}_{\gamma\gamma }  - 0.56\, \delta y_t \,,\\
&\frac{\Gamma_{\text{Z}\gamma}}{\Gamma_{Z\gamma}^{\text{SM}}} &= & 
\;1 +2.11\,\delta c_z- 3.4\, \hat{c}_{z\gamma }  - 0.113\, \delta y_t\,,\\
&\frac{\Gamma_{\text{WW}}}{\Gamma_{\text{WW}}^{\text{SM}}} &=&
\;1 +2.0\,\delta c_z   + 0.67\, c_{z\square } +0.05\, c_{zz} -0.0182\, \hat{c}_{z\gamma } -0.0051\, \hat{c}_{\gamma\gamma}\,,\\
&\frac{\Gamma_{\text{ZZ}}}{\Gamma_{\text{ZZ}}^{\text{SM}}} &=&
\;1 +2.0\,\delta c_z + 0.33\, c_{z\square } + 0.19\, c_{zz} -0.0081\, \hat{c}_{z\gamma } - 0.00111\, \hat{c}_{\gamma\gamma}\,,\\
&\frac{\Gamma_{\tau\tau}}{\Gamma_{\tau\tau}^{\text{SM}}} &=&
\;1 +2.0\,\delta y_\tau\,,\\
&\frac{\Gamma_{bb}}{\Gamma_{bb}^{\text{SM}}} &=&
\;1 +2.0\,\delta y_b\,,\\
&\frac{\Gamma_{\text{H}}}{\Gamma_{\text{H}}^{\text{SM}}} &=&
\;1+0.171\, \hat{c}_{gg}+0.006\, c_{zz}-0.0091\, \hat{c}_{z\gamma}+
0.15\, c_{z\square}-0.0061\, \hat{c}_{\gamma \gamma}+0.48\, \delta c_{z} \nn\\
	& & &+ 1.15\, \delta y_{b}+0.23\, \delta y_t +0.13\, \delta y_{\tau}\,,
\end{alignat}
where in the modification of the decay to two photons we made use of the one-loop result\footnote{We observed that the NLO corrections in the $\gamma\gamma$ decay have no impact on the global fit once enough observables are included to remove the flat directions.} of ref.~\cite{Hartmann:2015aia}, suitably translated to the Higgs basis and evaluated at the renormalization scale $\mu = m_h$. The analog result for the decay to $Z\gamma$ is not yet available in the literature, and we only include the known terms. In any case, the corresponding branching ratio will be measured with a limited precision and the impact of the missing one-loop corrections is going to be negligible.

For completeness we also report the expressions for the dependence of the Higgs rates on the modification of the
Higgs self-coupling $\kappa_\lambda$. These results were derived in ref.~\cite{Degrassi:2016wml}.
The modification to the Higgs production and decay rates can be parametrized as
\begin{equation}\label{eq:sigma_NLO}
\frac{\sigma}{\sigma_\text{SM}}=1+(\kappa_{\lambda}-1) C^\sigma +  \frac{(\kappa_\lambda^2-1)\delta Z_H}{1-\kappa_\lambda^2 \delta Z_H}\,,
\end{equation}
and
\begin{equation}\label{eq:gamma_NLO}
\frac{\Gamma}{\Gamma_\text{SM}}=1+(\kappa_{\lambda}-1) C^\Gamma +  \frac{(\kappa_\lambda^2-1)\delta Z_H}{1-\kappa_\lambda^2 \delta Z_H}\,.
\end{equation}
In the above expressions the term linear in $\kappa_\lambda$ comes from diagrams that contribute directly to the production
and decay processes. The corresponding coefficients $C^{\sigma}$ and $C^\Gamma$ for the inclusive cross sections
are given in tables~\ref{tab:cdecay} and \ref{tab:cprod}.
The last terms in eqs.~(\ref{eq:sigma_NLO}) and (\ref{eq:gamma_NLO}) comes from a rescaling of the Higgs kinetic term due to
the self-energy diagram involving two insertions of the Higgs self-coupling. The corresponding quantity $\delta Z_H$ is given by
\begin{equation}
\delta Z_H = -\frac{9}{16}\frac{G_\mu m_H^2}{\sqrt{2}\pi^2}\left(\frac{2\pi}{3\sqrt{3}}-1\right) \simeq - 0.0015\,.
\end{equation}
\begin{table}
	\begin{center}
			\bgroup
			\def\arraystretch{1.1}
			\begin{tabular}{l|c|c|c|c|c}
				$C^\Gamma\ \ [\%]$ &$ \gamma \gamma$ &  $ZZ$ & $WW$ & $f \bar{f}$ & $gg$ \\  \hline \hline 
				H & 0.49 &0.83&0.73&0&0.66 
			\end{tabular}
			\egroup
		\caption{Coefficients parametrizing the corrections to the Higgs partial widths due to loops involving the Higgs self-coupling
		(see eq.~(\ref{eq:gamma_NLO}))~\cite{Degrassi:2016wml}.}\label{tab:cdecay}
	\end{center}
\end{table}
\begin{table}
	\begin{center}
		\bgroup
		\def\arraystretch{1.1}
		\begin{tabular}{r|c|c|c|c|c}
			\multicolumn{1}{c|}{$C^\sigma\ \ [\%]$} & ggF &  VBF & $WH$ & $ZH$ & $t \overline{t} H$ \\  \hline \hline 
			7 $\tev{}$  & 0.66 &0.65	&1.06&1.23	&3.87  \\
			8 $\tev{}$  & 0.66 &0.65	&1.05&1.22	&3.78 \\
			13 $\tev{}$ & 0.66 &0.64	&1.03&1.19	&3.51  \\
			14 $\tev{}$ & 0.66 &0.64	&1.03&1.18	&3.47 
		\end{tabular}
		\egroup
		 \caption{Coefficients parametrizing the corrections to the Higgs production cross sections
		 due to loops involving the Higgs self-coupling (see eq.~(\ref{eq:sigma_NLO}))~\cite{Degrassi:2016wml}.}\label{tab:cprod}
	\end{center}
\end{table}

\medskip

We now report the expressions for the Higgs pair production differential cross section.
This cross-section has been calculated in the EFT framework in ref.~\cite{Azatov:2015oxa},
as a function of the parameters $\delta y_t$, $\delta y_t^{(2)}$, $\hat{c}_{gg}$,
$\hat{c}_{gg}^{(2)}$, and $\kappa_\lambda$.
The ratio of the inclusive cross-section for Higgs-pair production to the corresponding SM
prediction can be written as
\begin{align}
  \frac{\sigma(pp\to hh)}{\sigma_{\textsc{sm}}(pp\to hh)} = {} &
  A_1 \, (1+\delta y_t)^4
+ A_2 \, (\delta y_t^{(2)})^2
 + A_3 \, \kappa_\lambda^2\, (1+\delta y_t)^2
+ A_4 \, \kappa_\lambda^2\,\hat{c}_{gg}^2\nn\\
&   + A_5 \, (\hat{c}_{gg}^{(2)})^2
     + A_6 \, (1+\delta y_t)^2\, \delta y_t^{(2)}
+ A_7 \, \kappa_\lambda\, (1+\delta y_t)^3\nn\\
& + A_8 \, \kappa_\lambda\, (1+\delta y_t)\, \delta y_t^{(2)}
+ A_9 \, \kappa_\lambda\, \hat{c}_{gg}\, \delta y_t^{(2)}
 + A_{10} \, \hat{c}_{gg}^{(2)}\, \delta y_t^{(2)}\nn\\
& + A_{11} \, \kappa_\lambda\, \hat{c}_{gg}\, (1+\delta y_t)^2
+ A_{12} \, \hat{c}_{gg}^{(2)}\, (1+\delta y_t)^2
 + A_{13} \, \kappa_\lambda^2\, \hat{c}_{gg}\, (1+\delta y_t)\nn\\
 & + A_{14} \, \kappa_\lambda\, \hat{c}_{gg}^{(2)}\, (1+\delta y_t)
 + A_{15} \, \kappa_\lambda\, \hat{c}_{gg}\, \hat{c}_{gg}^{(2)} \,,
\label{eq:xhhnewnorm}
\end{align}
Notice that this parametrization can be used for the full uncut cross
section and also for the cross section obtained after imposing cuts
and acceptance factors. Moreover we can use the same expression to
parametrize the differential cross section in each bin of the
Higgs-pair invariant mass distribution.  We report in
table~\ref{tab:xshhcoeffs} the inclusive and differential SM cross
section at 14 \tev{} after imposing the cuts devised in
ref.~\cite{Azatov:2015oxa}, as well as the values of the $A_i$.

\begin{table}
\centering
\bgroup
\def\arraystretch{1.1}
  \begin{tabular}{c|c|cccccc}
    $m_{hh}^{\text{reco}}\,[{\rm GeV}]$ & \text{inclusive} & {250--400} & {400--550} & {550--700} & {700--850} & {850--1000} & {1000--} \\
    \hline
    \hline
    $\sigma _{\textsc{sm}}\ [{\rm ab}] $ & 1.6 & 0.27 & 0.8 & 0.36 & 0.13 & 0.042 & 0.021 \\
 $A_1$ & 1.7 & 2.3 & 1.7 & 1.5 & 1.3 & 1.2 & 1.2 \\
 $A_2$ & 2.7 & 1.8 & 2.1 & 3.2 & 4.7 & 6.4 & 9.1 \\
 $A_3$ & 0.12 & 0.27 & 0.11 & 0.057 & 0.034 & 0.022 & 0.011 \\
 $A_4$ & 0.042 & 0.094 & 0.037 & 0.026 & 0.024 & 0.023 & 0.022 \\
 $A_5$ & 1.5 & 0.62 & 0.69 & 1.5 & 3.5 & 7.1 & 20. \\
 $A_6$ & -3.8 & -4.0 & -3.6 & -3.8 & -4.2 & -4.5 & -4.6 \\
 $A_7$ & -0.82 & -1.5 & -0.84 & -0.51 & -0.36 & -0.26 & -0.17 \\
 $A_8$ & 0.98 & 1.4 & 0.96 & 0.83 & 0.78 & 0.73 & 0.67 \\
 $A_9$ & 0.45 & 0.81 & 0.46 & 0.33 & 0.23 & 0.14 & 0.003 \\
 $A_{10}$ & 2.2 & 2.1 & 2.0 & 2.4 & 2.8 & 2.5 & -0.56 \\
 $A_{11}$ & -0.32 & -0.88 & -0.33 & -0.081 & 0.03 & 0.087 & 0.13 \\
 $A_{12}$ & -1.0 & -2.3 & -1.3 & -0.6 & 0.33 & 1.6 & 4.1 \\
 $A_{13}$ & 0.12 & 0.33 & 0.11 & 0.044 & 0.02 & 0.0092 & 0.0014 \\
 $A_{14}$ & 0.46 & 0.82 & 0.44 & 0.36 & 0.29 & 0.13 & -0.27 \\
 $A_{15}$ & 0.41 & 0.48 & 0.31 & 0.39 & 0.57 & 0.81 & 1.3
  \end{tabular}
  \egroup
\caption{Coefficients parametrizing the inclusive and differential cross section for double Higgs production
via gluon fusion at $\sqrt{s}=14$ \tev{}. By $\sigma_{\textsc{sm}}$ we denote the SM cross section, while $A_1$--$A_{15}$ are the coefficients
parametrizing the dependence of the cross on the EFT parameters as defined in eq.~(\ref{eq:xhhnewnorm}).
The numerical results correspond to the ones derived in the analyses of ref.~\cite{Azatov:2015oxa}.}
\label{tab:xshhcoeffs}
\end{table}

\medskip

\begin{table}
  \centering
  \scalebox{.735}{
    \renewcommand*{\arraystretch}{1.2}
    \begin{tabular}{cc|rrrrrr|rrrrrr}
      \multirow{2}*{$\sqrt{s}$} & \multirow{2}*{$\sqrt{\hat{s}}/m_{\text{threshold}}$} & \multicolumn{6}{c|}{$WH$} & \multicolumn{6}{c}{$ZH$} \\
                                & & $\epsilon_{\text{SM}}$ & $\delta c_z$ & $c_{z {\scriptscriptstyle\square}}$ & $c_{zz}$ & $ \hat{c}_{z\gamma}$ & $\hat{c}_{\gamma \gamma}$ & $\epsilon_{\text{SM}}$ & $\delta c_z$ & $c_{z {\scriptscriptstyle\square}}$ & $c_{zz}$ & $ \hat{c}_{z\gamma}$ & $\hat{c}_{\gamma \gamma}$  \\
      \hline
      \hline
      \multirow{5}*{7 \tev}
                                & $[1.0-1.1]$ & 19 \% & 1.99 & 4.95 & 2.68 & -0.0270 & -0.0215 & 20 \% & 2.00 & 4.14 & 2.14 & -0.0220 & -0.0123 \\
                                & $[1.1-1.2]$ & 20 \% & 2.00& 5.84 & 3.10 & -0.0349 & -0.0258 & 21 \% & 2.00 & 4.81 & 2.42 & -0.0290 & -0.0154 \\
                                & $[1.2-1.5]$ & 35 \% & 2.00& 7.40 & 3.80 & -0.0504 & -0.0334 & 34 \% & 2.01 & 6.44 & 3.07 & -0.0447 & -0.0226 \\
                                & $[1.5-2.0]$ & 18 \% & 2.01 & 12.4 & 5.71 & -0.116 & -0.0598 & 17 \% & 2.01 & 10.5 & 4.44 &-0.0853 &-0.0393 \\
                                & $[2.0-3.0]$ & 7 \% & 2.01 & 23. & 9.38 & -0.271 & -0.117 & 6 \% & 1.98 & 19.7 & 6.90 & -0.192 & -0.0780 \\
      \hline
      \multirow{5}*{8 \tev}
                                & $[1.0-1.1]$ & 19 \% & 2.01 & 4.93 & 2.66 & -0.0275 & -0.0215 & 20 \% & 2.00 & 4.10 & 2.12 & -0.0231 & -0.0126 \\
                                & $[1.1-1.2]$ & 20 \% & 1.97 & 5.73 & 3.05 & -0.0337 & -0.0252 & 20 \% & 2.01 & 4.90 & 2.49 & -0.0299 & -0.0158 \\
                                & $[1.2-1.5]$ & 34 \% & 2.01 & 7.51 & 3.81 & -0.0533 & -0.0342 & 35 \% & 2.01 & 6.40 & 3.05 & -0.0453 & -0.0226 \\
                                & $[1.5-2.0]$ & 19 \% & 1.99 & 12.1 & 5.56 & -0.113 & -0.0582 & 18 \% & 2.00 & 10.6 & 4.51 & -0.0872 & -0.0400 \\
                                & $[2.0-3.0]$ & 7 \% & 2.02 & 22.3 & 9.12 & -0.264 & -0.114 & 6 \% & 1.95 & 20.0 & 6.99 & -0.202 & -0.0804 \\
      \hline
      \multirow{5}*{13 \tev}
                                & $[1.0-1.1]$ & 18 \% & 2.02 & 4.96 & 2.70 & -0.0265 & -0.0216 & 19 \% & 2.02 & 4.06 & 2.09 & -0.0226 & -0.0121 \\
                                & $[1.1-1.2]$ & 19 \% & 1.97 & 5.81 & 3.08 & -0.0344 & -0.0256 & 20 \% & 2.00& 4.86 & 2.45 & -0.0300 & -0.0157 \\
                                & $[1.2-1.5]$ & 34 \% & 2.00& 7.44 & 3.76 & -0.0532 & -0.0339 & 34 \% & 1.98 & 6.37 & 3.04 & -0.0445 & -0.0222 \\
                                & $[1.5-2.0]$ & 19 \% & 2.02 & 11.9 & 5.46 & -0.111 & -0.0572 & 18 \% & 2.01 & 10.6 & 4.53 & -0.0887 & -0.0406 \\
                                & $[2.0-3.0]$ & 8 \% & 1.99 & 22.6 & 9.20 & -0.269 & -0.116 & 7 \% & 2.00& 20.4 & 7.29 & -0.196 & -0.0808 \\
      \hline
      \multirow{5}*{14 \tev}
                                & $[1.0-1.1]$ & 18 \% & 2.00& 5.01 & 2.72 & -0.0267 & -0.0217 & 19 \% & 2.01 & 4.14 & 2.12 & -0.0237 & -0.0126 \\
                                & $[1.1-1.2]$ & 19 \% & 2.00& 5.81 & 3.10 & -0.0337 & -0.0255 & 20 \% & 2.01 & 4.86 & 2.49 & -0.0284 & -0.0156 \\
                                & $[1.2-1.5]$ & 34 \% & 2.01 & 7.44 & 3.76 & -0.0535 & -0.0340 & 34 \% & 2.00& 6.35 & 3.02 & -0.0448 & -0.0221 \\
                                & $[1.5-2.0]$ & 19 \% & 1.98 & 11.8 & 5.40 & -0.112 & -0.0572 & 18 \% & 1.98 & 10.5 & 4.44 & -0.0873 & -0.0396 \\
                                & $[2.0-3.0]$ & 8 \% & 2.03 & 22.6 & 9.05 & -0.276 & -0.117 & 7 \% & 1.96 & 20.3 & 7.27 & -0.193 & -0.0800 \\
      \hline
      \multirow{5}*{33 \tev}
                                & $[1.0-1.1]$ & 17 \% & 1.98 & 4.96 & 2.68 & -0.0274 & -0.0216 & 18 \% & 2.02 & 4.16 & 2.16 & -0.0228 & -0.0124 \\
                                & $[1.1-1.2]$ & 18 \% & 2.01 & 5.77 & 3.07 & -0.0338 & -0.0254 & 19 \% & 1.99 & 4.77 & 2.41 & -0.0282 & -0.0150 \\
                                & $[1.2-1.5]$ & 33 \% & 1.99 & 7.43 & 3.73 & -0.0544 & -0.0340 & 34 \% & 1.99 & 6.45 & 3.08 & -0.0453 & -0.0225 \\
                                & $[1.5-2.0]$ & 20 \% & 2.00& 12.00& 5.54 & -0.110 & -0.0574 & 19 \% & 2.02 & 10.4 & 4.37 & -0.0862 & -0.0390 \\
                                & $[2.0-3.0]$ & 9 \% & 2.02 & 23.3 & 9.56 & -0.274 & -0.119 & 8 \% & 2.00& 19.8 & 6.97 & -0.190 & -0.0777 \\
      \hline
      \multirow{5}*{100 \tev}
                                & $[1.0-1.1]$ & 16 \% & 2.01 & 4.92 & 2.66 & -0.0271 & -0.0215 & 17 \% & 2.02 & 3.98 & 2.05 & -0.0238 & -0.0118 \\
                                & $[1.1-1.2]$ & 18 \% & 2.04 & 5.82 & 3.09 & -0.0344 & -0.0257 & 18 \% & 2.00& 5.02 & 2.60 & -0.0282 & -0.0157 \\
                                & $[1.2-1.5]$ & 33 \% & 1.97 & 7.48 & 3.77 & -0.054 & -0.0341 & 33 \% & 2.00& 6.45 & 3.09 & -0.0445 & -0.0224 \\
                                & $[1.5-2.0]$ & 20 \% & 2.02 & 11.9 & 5.47 & -0.111 & -0.0573 & 20 \% & 1.99 & 10.5 & 4.38 & -0.0860 & -0.0389 \\
                                & $[2.0-3.0]$ & 10 \% & 1.99 & 23.1 & 9.40 & -0.275 & -0.118 & 9 \% & 2.00& 20.0 & 6.90 & -0.195 & -0.0782
    \end{tabular}
  }
  		\caption{Effective field theory coefficient for each bin in the $\hat s$ differential distribution. The
  		bins extrema are expressed in units of $m_{\text{threshold}} \equiv m_V + m_h$.
  		The $\epsilon_{\textsc{sm}}$ columns list the percentage of events that belong to each bin in the SM
  		distribution.}\label{tab:dim6diff}
\end{table}

Finally we consider the differential distributions for the Higgs associated production channels.
In table~\ref{tab:dim6diff} we list the dependence of the differential cross section in $ZH$ and $WH$
on the single-Higgs EFT parameters.
The results are presented for the binned invariant mass $\hat s$ distribution.
The cross sections have been computed by analyzing the events generated at LO by~\texttt{MadGraph}
through a custom made C++ code based on the~\texttt{MadAnalysis5} library~\cite{Conte:2014zja,Conte:2012fm}.
The coefficients that parametrize the dependence of the $WH$, $ZH$ and $t\overline t H$ production channels
on the Higgs self-coupling are listed in table~\ref{tab:cmtdiff}.

In tables~\ref{tab:differential2_error} and~\ref{tab:differential_error} we list the estimates of the systematic uncertainties
on the binned differential distributions. To estimate the expected errors on the inclusive cross sections,
we compared the ATLAS projections for the $300/$fb and $3/$ab experimental uncertainties
and assumed that they come from a sum in quadrature of systematic and statistical ones.
In the `optimistic' scenario in table~\ref{tab:differential2_error}, we rescaled the statistical uncertainty by the square
root of the ratio of SM number of events in each bin, whereas we kept the systematic errors constant. In the `pessimistic' scenario we rescaled the total (statistical plus systematic) uncertainty according to the
number of events in each bin.

\begin{table}
	\begin{center}
		\resizebox{0.9\columnwidth}{!}{
		\bgroup
		\def\arraystretch{1.1}
			\begin{tabular}{c|r|r|r|r|r}
				$C^{\sigma}\ \ [\%]$&\multicolumn{1}{c|}{$[1.0-1.1]$}&\multicolumn{1}{c|}{$[1.1-1.2]$}&\multicolumn{1}{c|}{$[1.2-1.5]$}&\multicolumn{1}{c|}{$[1.5-2.0]$}&\multicolumn{1}{c}{$[2.0-3.0]$} \\ \hline \hline
				${WH}$&1.78 (0.18)&1.44 (0.19)&1.02 (0.34)&0.52 (0.19)&0.06 (0.08)  \\ 
				${ZH}$&2.08 (0.19)&1.64 (0.20)&1.12 (0.34)&0.51 (0.18)&0.21 (0.07)  \\ 
				${t \overline{t} H}$&8.57 (0.02)&6.63 (0.08)&4.53 (0.33)&2.83 (0.33)&1.61 (0.18)
			\end{tabular}
					\egroup
		}
		 \caption{Coefficients parametrizing the corrections to the differential Higgs production cross sections at $13$~TeV in the
		         $WH$, $ZH$ and $t \overline t H$ channels
		 		 due to loops involving the Higgs self-coupling (see eq.~(\ref{eq:sigma_NLO})).
		 		 The bins extrema are expressed in units of $m_{\text{threshold}}$, defined as $m_{\text{threshold}} \equiv m_V + m_h$ for $WH$ and $ZH$, and $m_{\text{threshold}} \equiv 2 m_t + m_h$ for $t \overline t H$.
		 		 In parentheses we give the fraction of events belonging to each bin in the SM distribution.
		 		 The results are taken from ref.~\cite{Degrassi:2016wml}.}\label{tab:cmtdiff}
	\end{center}
\end{table}

\begin{table*}
	\begin{center}
		\begin{tabular}{ll|c|c@{\hspace{.5em}}c@{\hspace{.5em}}c@{\hspace{.5em}}c@{\hspace{.5em}}c}
			\multicolumn{2}{c|}{Process} & Systematic & $[1.0-1.1]$ & $[1.1-1.2]$ & $[1.2-1.5]$ & $[1.5-2.0]$ & $[2.0-3.0]$ \\ \hline \hline
			\multirow{3}{*}{${H} \to \gamma \gamma$}&${t\overline tH}$  & 0.04 & 0.74 & 0.41 & 0.23 & 0.23 & 0.3 \\ 
			&${WH}$& 0.08 & 0.37 & 0.37 & 0.28 & 0.36 & 0.54 \\
			&${ZH}$ & 0.03 & 0.62 & 0.61 & 0.47 & 0.63 & 0.99 \\ \hline
			\multirow{3}{*}{${H}\to {ZZ}$}&${t\overline tH}$ & 0.05 & 0.98 & 0.53 & 0.29 & 0.29 & 0.39 \\ 
			&${WH}$& 0.07 & 0.33 & 0.32 & 0.25 & 0.32 & 0.48 \\
			&${ZH}$ & 0.09 & 0.42 & 0.41 & 0.32 & 0.42 & 0.65 \\\hline
			\multirow{2}{*}{${H}\to {b\bar b}$} &${WH}$ & 0.33 & 0.48 & 0.48 & 0.42 & 0.47 & 0.61 \\ 
			&${ZH}$ & 0.10 & 0.23 & 0.22 & 0.18 & 0.23 & 0.34
		\end{tabular}
		\caption{Estimated relative uncertainties on the determination of the differential distributions in the associated Higgs production
		channels. These estimates correspond to the `optimistic' scenario in which the systematic uncertainties are assumed
		to be the same for each bin and only the statistical uncertainty is rescaled according to the number of events in each bin.}
		\label{tab:differential2_error}
	\end{center}
\end{table*}

\section{Trilinear gauge couplings}\label{app:trilinear}

The deviations in the EW boson trilinear gauge couplings induced by CP-preserving dimension-6 operators can be encoded in
the following effective Lagrangian
\begin{align}\nn
\mathcal{L} \supset &\; i\, \gL{}\, c_{w}\, \delta g_{1,z} \left(W_{\mu \nu}^+ W^{\mu-}-W_{\mu \nu}^- W^{\mu+}\right)Z^{\nu}\\\nn
&+ i\, e\,\delta \kappa_{\gamma} \,A^{\mu\nu}W_{\nu}^+W_{\nu}^- + i\, \gL{}\, c_w\, \delta \kappa_z\, Z^{\mu\nu} W_{\mu}^+W_{\nu}^- \\
&+ i\, \frac{e\, \lambda_\gamma}{m_w^2} W_{\ \; \nu}^{\mu+} W_{\ \; \rho}^{\nu-}A^{\rho}_{\ \mu} + \frac{g\, c_{w}\, \lambda_Z}{m_w^2} W_{\ \; \nu}^{\mu+} W_{\ \; \rho}^{\nu-}Z^{\rho}_{\ \mu}\,,
\end{align}
where $s_w$ and $c_w$ denote the sine and cosine of the weak mixing angle.

In the Higgs basis the above couplings depend only on one free parameter, $\lambda_z$, while the other coefficients are combinations
of the Higgs coupling parameters $\hat c_{\gamma\gamma}$, $\hat c_{z\gamma}$, $c_{zz}$ and $c_{z\square}$.
The explicit expressions are given by
\begin{align}
\delta g_{1,z}={}&\frac{\gY{}^2}{2(\gL{}^2-\gY{}^2)}\left[ \vphantom{\frac{\gL{}^2}{\gY{}^2}} \hat{c}_{\gamma\gamma} \frac{e^2}{\pi^2}+ \hat{c}_{z\gamma}\frac{\gL{}^2-\gY{}^2}{\pi^2} - \right.  \nonumber\\
& \qquad \qquad \quad \;\left.c_{zz}\left(\gL{}^2+\gY{}^2\right) -c_{z\square}\frac{\gL{}^2}{\gY{}^2}\left(\gL{}^2+\gY{}^2\right)\right]\,, \label{eq:tgc_g1z}\\
\delta\kappa_\gamma ={}& -\frac{\gL{}^2}{2 (\gL{}^2 + \gY{}^2)}\left[\hat{c}_{\gamma\gamma} \frac{e^2}{\pi^2} +  \hat{c}_{z\gamma}\frac{\gL{}^2-\gY{}^2}{\pi^2} - c_{zz}(\gL{}^2 + \gY{}^2)\right], \label{eq:tgc_kappa_gamma}\\
\delta\kappa_z={}&\delta g_{1,z}-\frac{\gY{}^2}{\gL{}^2}\delta \kappa_{\gamma}\,,\\
\lambda_\gamma ={}& \lambda_z\,.
\end{align}

\begin{table*}
	\begin{center}
		\begin{tabular}{ll|ccccc}
			\multicolumn{2}{c|}{Process} & $[1.0-1.1]$ & $[1.1-1.2]$ & $[1.5-1.2]$ & $[2.0-1.5]$ & $[2.0-3.0]$ \\ \hline \hline
			\multirow{3}{*}{${H} \to \gamma \gamma$}&${t\overline tH}$  & 0.78 & 0.43 & 0.24 & 0.24 & 0.31 \\
			&${WH}$  & 0.41 & 0.4 & 0.3 & 0.4 & 0.6 \\
			&${ZH}$  & 0.63 & 0.62 & 0.47 & 0.63 & 0.99 \\ \hline
			\multirow{3}{*}{${H}\to {ZZ}$}&${t\overline tH}$ & 1.04 & 0.56 & 0.3 & 0.3 & 0.4 \\
			&${WH}$ & 0.37 & 0.36 & 0.27 & 0.35 & 0.53 \\
			&${ZH}$ & 0.46 & 0.45 & 0.35 & 0.47 & 0.72 \\\hline
			\multirow{2}{*}{${H}\to {b\bar b}$}&${WH}$ & 0.86 & 0.84 & 0.62 & 0.82 & 1.26 \\
			&${ZH}$ & 0.3 & 0.3 & 0.23 & 0.31 & 0.48
		\end{tabular}
		\caption{Estimated relative uncertainties on the determination of the differential distributions in the associated Higgs production
				channels. These estimates correspond to the `pessimistic' scenario in which the total (statistical plus systematic)
				uncertainty is rescaled according to the number of events in each bin.}
		\label{tab:differential_error}
	\end{center}
\end{table*}

\vspace{1cm}

\providecommand{\href}[2]{#2}\begingroup\raggedright\endgroup

\end{document}